\documentclass[final,5p,times,twocolumn,authoryear]{elsarticle}

\usepackage{amssymb}
\usepackage{amsmath}
\usepackage{mathtools}
\usepackage{algorithm}
\usepackage{algpseudocode}
\usepackage{textcomp}
\usepackage{hyperref}
\hypersetup{colorlinks=true, allcolors=blue}

\journal{International Journal of Rock Mechanics and Mining Sciences}

\begin{document}
	
	\begin{frontmatter}
		
		\title{Intelligent Prediction and Optimization of Open-Hole Wellbore Multiphysics Stability: A Synergistic PINN-DRL Approach}
		
		\author[a]{Yu Song\corref{cor1}}
		\ead{songyu_cup@163.com}
		\cortext[cor1]{Corresponding author.}
		
		\author[a]{Zehua Song}
		\author[b]{Jin Yang}
		\author[b]{Kejin Chen}
		\author[b]{Kun Jiang}
		\author[c]{Jizhou Tang}
		
		\affiliation[a]{organization={College of Artificial Intelligence, China University of Petroleum (Beijing)},
			city={Beijing},
			postcode={102249}, 
			country={China}}
		
		\affiliation[b]{organization={College of Safety and Ocean Engineering, China University of Petroleum (Beijing)},
			city={Beijing},
			postcode={102249}, 
			country={China}}
		
		\affiliation[c]{organization={School of Ocean and Earth Science, Tongji University},
			city={Shanghai},
			postcode={200092}, 
			country={China}}
		
		\begin{abstract}
			To address the dual challenge of predicting multiphysics-induced instability and optimizing drilling fluid parameters for open-hole wellbores under long-term exposure, a high-fidelity system of coupled governing equations was developed. This system integrates seepage, hydration-induced softening, thermal diffusion, and elasto-plastic response to capture the nonlinear dynamics of wellbore stability evolution. A two-dimensional numerical model in a polar coordinate system was established using COMSOL Multiphysics to simulate multi-lithology and multi-parameter perturbations. This process generated a high-dimensional dataset characterizing the evolution of Von Mises stress, plastic strain, pore pressure, temperature, and water content, and its physical consistency was examined. Subsequently, the Seepage–Thermal–Water–Mechanical Physics-Informed Neural Network (STWM-PINN) is proposed. This model embeds governing equation residuals and initial-boundary constraints to achieve high-precision, physically consistent predictions of the wellbore’s spatio-temporal evolution under the supervision of finite observational data, laying a foundation for parameter control. Building on this, a Double-Noise Soft Actor-Critic (DN-SAC) algorithm is integrated. A reward function was designed to minimize the probability of instability while considering control smoothness and physical boundary constraints, enabling continuous-space optimization of drilling fluid parameters. A case study demonstrates that the proposed method delays the onset of instability by an average of 32.33\% and a maximum of 53.35\%, significantly reducing instability risk. This study provides a decision-support framework with engineering application potential for intelligent wellbore instability prediction and drilling fluid control.
		\end{abstract}
		
		
	\end{frontmatter}
	
	\section{Introduction}
	\label{sec:intro}
	
	Driven by the global energy transition and carbon-neutrality strategies, the development of offshore oil and gas fields has become a key direction for improving economic and operational efficiency \citep{ref1, ref2}. The multiphysics coupling of seepage, thermal, hydration, and stress effects, induced by dynamic drilling fluid invasion, significantly increases the risk of progressive wellbore instability \citep{ref3, ref4}. Statistics from the Bohai Sea and surrounding regions indicate that wellbore instability incidents can increase non-productive time by up to 23\%. This not only leads to single-incident costs exceeding 20 million USD but, more importantly, significantly delays the production of critical energy resources, impacting regional energy supply security \citep{ref5}. Therefore, an intelligent warning and control system is urgently needed. Such a system must be capable of efficiently and accurately predicting the evolution of open-hole stability while supporting the intelligent optimization of drilling fluid parameters and critical casing-running time decisions \citep{ref6, ref7, ref8}. This is essential to ensure the continuity, safety, and economic viability of offshore drilling operations.
	
	Traditional wellbore stability assessment has long relied on static safety factor methods based on elasto-plastic finite element methods and the Mohr–Coulomb criterion, a practice included in the API RP 13B-1 standard \citep{ref9}. However, these methods fail to consider the time-dependent effect of pore pressure redistribution from dynamic drilling fluid seepage. Their predictive accuracy is limited in formations with well-developed bedding and strong heterogeneity, often leading to significant deviations. Recent research has sought to overcome these limitations. In high-fidelity numerical simulation, Pirhadi et al. developed a coupled thermo-poro-elastic finite element model, revealing how drilling fluid temperature regulates tensile and shear failure modes in depleted reservoirs \citep{ref10}. However, their model did not account for the superposition effect of mechanical weakening due to hydration, limiting its applicability to shale formations. Ding et al. established a coupled stress-unloading and hydration model to quantify the combined damage from bedding plane weakening and fluid invasion, but they did not resolve the time-dependent shielding effect of the mud-cake deposition process \citep{ref11}. Saber et al. optimized horizontal well orientation to balance productivity and stability using a COMSOL dual-porosity model, but its reliance on inverted reservoir parameters limited its generalization capability \citep{ref12}. Feng et al. proposed a dynamic mud-cake model describing thickness and permeability effects on the stress field, yet it did not fully capture seepage–stress feedback, leading to inaccurate long-term predictions \citep{ref13}. Similarly, in hydraulic fracturing, some studies use high-precision numerical methods like the discrete lattice method for detailed simulations of fracture initiation and extension. These have revealed the complex influence of perforation parameters on fracture geometry \citep{ref14}, but the computational cost of such methods limits their application in optimization scenarios requiring rapid decisions. In data-driven surrogate modeling, Wu et al. used artificial neural networks (ANNs) trained on offset well data to predict stability, thereby reducing dependence on expert judgment \citep{ref15}. However, the adaptability of the model to new geological blocks was insufficient due to its limited training domain. Udegbunam et al. applied Monte Carlo methods to quantify the impact of input parameter uncertainty on collapse pressure but neglected parameter correlations and error accumulation \citep{ref16}. Chen et al. introduced the Nataf transform and Kolmogorov–Smirnov test to build a risk analysis framework for non-independent parameters, significantly narrowing the uncertainty range but increasing model complexity and limiting its real-time potential \citep{ref17}. Rafieepour et al. proposed a simplified chemo-thermo-poro-elastic model for improved engineering usability but did not embed an anisotropic damage mechanism, making it difficult to accurately capture early-stage instability \citep{ref18}. In recent years, deep learning techniques such as Convolutional Neural Networks (CNNs) have also shown great potential for related tasks, including intelligent lithology recognition from geological core images \citep{ref19}. Concurrently, intelligent modeling methods for wellbore stability have gained increasing attention \citep{ref20, ref21}. However, they share two common limitations. First, high-fidelity multiphysics models can capture instability mechanisms in detail but are too computationally expensive for the minute-level response times required for real-time drilling fluid optimization. Second, while data-driven methods offer high inference speed, their lack of physical constraints creates reliability issues in extrapolation and optimization for complex well sections. Therefore, developing a predictive model that maintains high computational efficiency while rigorously embedding core multiphysics coupling mechanisms is a key challenge facing the field.
	
	To address the need for high-precision prediction and active control of drilling fluid parameters for the “seepage-thermal-hydration-mechanics” coupled instability process in open-hole wellbores, a synergistic optimization framework combining a multiphysics neural network and deep reinforcement learning is designed and developed. This framework is centered on the Seepage–Thermal–Water–Mechanical Physics-Informed Neural Network (STWM-PINN). A multi-branch coupled network architecture is constructed, which embeds the Darcy’s flow equation, the non-steady-state heat transfer equation, the hydration diffusion equation, and the Drucker–Prager yield criterion into its loss function as strong-form residual constraints. Combined with initial and boundary condition errors, this approach drives the neural network to achieve physically consistent spatio-temporal evolution predictions under the supervision of finite observational data. A deep reinforcement learning optimization mechanism is further introduced. This mechanism uses the spatial statistical average features of the network’s outputs (pore pressure, temperature, water content, equivalent stress, and plastic strain) as state observations. A continuous and adjustable action space for drilling fluid density, viscosity, and temperature is constructed. By designing a triple-weighted reward function that integrates instability risk, control smoothness, and physical boundary penalties, a physics-constrained optimization objective is formed. Policy optimization is performed using the Double-Noise Soft Actor-Critic (DN-SAC) algorithm. This algorithm utilizes a dual-disturbance mechanism of policy reparameterization noise and external Gaussian exploration noise to enhance exploration efficiency in the high-dimensional parameter space while ensuring gradient differentiability. The process generates an optimization policy for drilling fluid parameters, effectively extending the stable exposure time of the open-hole section and suppressing instability risk. This provides a decision-support method with engineering application potential for intelligent drilling fluid optimization in complex offshore formations.
	
	\section{Workflow}
	\label{sec:workflow}
	
	\textbf{High-Fidelity Simulation of Multiphysics-Coupled Wellbore State:} The first stage examines the stability evolution of an open-hole wellbore under long-term exposure to multiphysics disturbances from seepage, stress, thermal, and hydration effects. A high-fidelity, two-dimensional (2D), plane-strain numerical model was constructed in a polar coordinate system. This model, implemented in COMSOL, facilitates a strongly coupled solution of Biot’s modified momentum conservation, Darcy’s flow, non-steady-state heat transfer, and hydration diffusion. The model embeds hydration-dependent degradation relationships for the elastic modulus, internal friction angle, and cohesion. It also uses the Drucker–Prager yield criterion to describe the nonlinear elasto-plastic response. Parameters were determined based on statistics from well logs and core experiments in the Caofeidian block. Stratified Latin Hypercube Sampling was used to generate input samples, considering regional lithology and thermal properties. Adaptive mesh refinement was employed to ensure computational accuracy in high-gradient zones. The simulation outputs included a spatio-temporal series of Von Mises stress, plastic strain, pore pressure, and water content. This process provides a dataset with strong physical constraints to support the physics-informed neural network.
	
	\textbf{Construction and Training of the Physics-Informed Neural Network:} The second stage involves building and training the STWM-PINN to achieve high-precision prediction of the wellbore state evolution under coupled multiphysics disturbances. The network takes spatio-temporal coordinates as input and outputs pore pressure, temperature, water content, equivalent stress, and plastic strain. Through automatic differentiation, the residuals of the governing equations were constructed. The residuals of the Darcy’s flow, heat transfer, and hydration diffusion equations, as well as the Drucker-Prager yield criterion, were incorporated into the loss function to enforce physical consistency. A weighted loss was constructed by combining these physics residuals with boundary condition and finite observational data errors, thereby balancing physical principles with data fitting. Training was performed with a staged cosine annealing learning rate schedule to enhance stable convergence and generalization capability. This stage lays a solid predictive foundation for the intelligent control of drilling fluid parameters.
	
\begin{figure*}[t]
		\centering
		\includegraphics[width=1\textwidth]{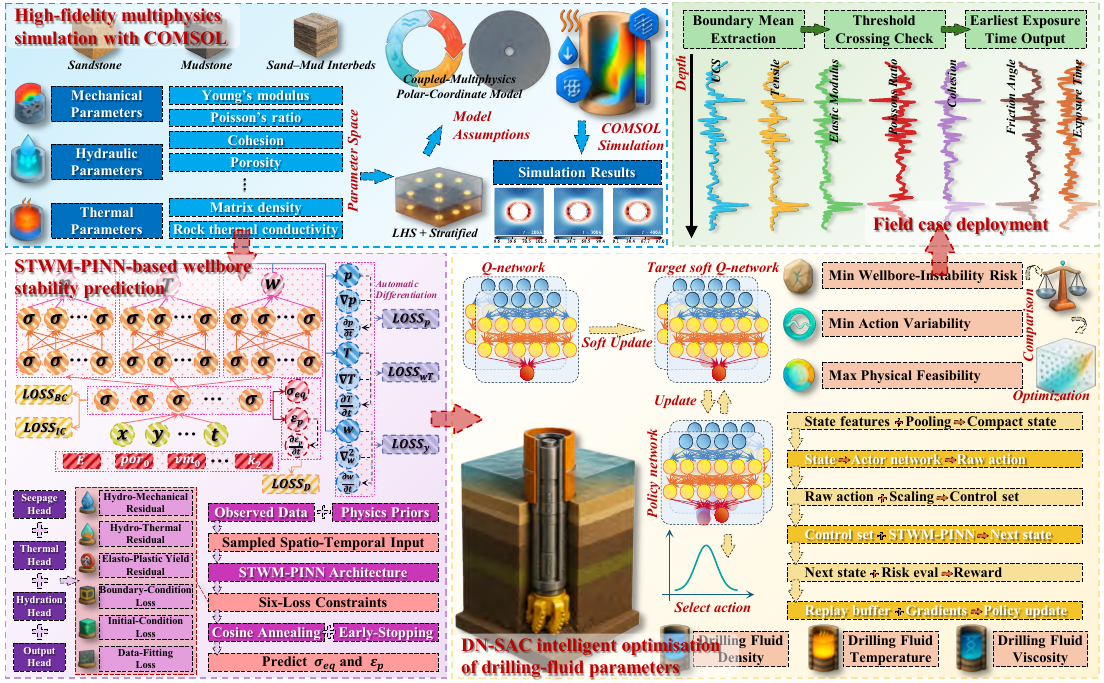} 
		\vspace*{-8mm} 
		\caption{Workflow.}
		\label{fig:workflow}
	\end{figure*}

	\textbf{Development of a Deep Reinforcement Learning-Based Parameter Optimization Policy:} The final stage builds on the STWM-PINN predictions to establish a deep reinforcement learning (DRL) framework for wellbore stability optimization. The state space is defined by the spatial statistical features of the PINN’s outputs, including pore pressure, temperature, water content, equivalent stress, and plastic strain. A continuous and adjustable action space is defined by the drilling fluid density, viscosity, and temperature. An immediate reward function was designed to combine penalties for instability risk, control smoothness, and physical boundary violations, thus creating an interpretable feedback mechanism. The Double-Noise Soft Actor-Critic (DN-SAC) algorithm was designed to enhance exploration capabilities and ensure policy differentiability through its dual-noise mechanism. This approach enables efficient, optimal decision-making for drilling fluid parameters, reducing the probability of failure and extending the stable exposure time. The overall workflow is illustrated in Fig.~\ref{fig:workflow}.
	
	\section{Multiphysics Coupled Model Simulation}
	\label{sec:simulation}
	
	\subsection{Coupled Governing Equations}
	\label{subsec:equations}
	
	To reveal the instability mechanism of open-hole wellbores under the coupled effects of seepage, thermal, hydration, and stress fields in complex geological settings, a comprehensive system of governing equations was established. This system integrates seepage-field disturbance, hydration-induced rock softening, formation temperature changes, and elasto-plastic mechanical response. The model is designed to fully describe the nonlinear evolution of wellbore stability influenced by factors such as drilling fluid invasion, heat exchange, and lithological variations. This provides a unified theoretical framework for subsequent high-fidelity numerical simulations and physics-constrained machine learning research.
	
	For the mechanical field, the governing equation for the rock mass behavior adopts the modified momentum conservation form of Biot’s theory \citep{ref22}. This is used to describe the stress equilibrium state in a consolidated porous medium. Its specific form is as follows:
	\begin{equation}
		\sigma_{ij,j}-(\alpha\delta_{ij} p)_{,j}+f_i=0
		\label{eq:1}
	\end{equation}
	where $\sigma_{ij}$ represents the effective stress tensor (MPa); $p$ represents the pore pressure (MPa); $\alpha$ represents the Biot coefficient (dimensionless); $\delta_{ij}$ represents the Kronecker delta (dimensionless), and $f_i$ represents the body force component (MPa).
	
	The local strain of the rock was decomposed according to incremental elasto-plasticity theory \citep{ref23}. Its constitutive relationship satisfied:
	\begin{equation}
		\begin{cases}
			d\varepsilon_{ij}=d\varepsilon_{ij}^e+d\varepsilon_{ij}^p+d\varepsilon_{ij}^T+d\varepsilon_{ij}^w \\
			d\sigma_{ij}=[D_e ]d\varepsilon_{ij}^e 
		\end{cases}
		\label{eq:2}
	\end{equation}
	where $d\varepsilon_{ij}$ represents the total strain increment (dimensionless); $d\varepsilon_{ij}^e$ represents the elastic strain increment (dimensionless); $d\varepsilon_{ij}^p$ represents the plastic strain increment (dimensionless); $d\varepsilon_{ij}^T$ represents the thermal expansion strain increment (dimensionless); $d\varepsilon_{ij}^w$ represents the hydration-induced swelling strain increment (dimensionless); $d\sigma_{ij}$ represents the effective stress increment (MPa); and $[D_e]$ represents the elastic stiffness matrix (MPa).
	
	The evolution of the plastic strain increment follows the associated flow rule, expressed mathematically as \citep{ref23}:
	\begin{equation}
		d\varepsilon_{ij}^p=d\lambda \frac{\partial g}{\partial\sigma_{ij}}
		\label{eq:3}
	\end{equation}
	where $g$ represents the plastic potential function (MPa) and $d\lambda$ represents the plastic multiplier increment (dimensionless). Under ideal plasticity conditions, $g$ can be set equal to the yield function $F$.
	
	To adapt to the characteristics of composite sand-shale formations, the model’s yield function employed the Drucker–Prager criterion \citep{ref24}. This criterion was used to determine the plastic failure state of the rock mass. Its expression is as follows:
	\begin{equation}
		F(I_1,J_2 )=aI_1+\sqrt{J_2}-k
		\label{eq:4}
	\end{equation}
	where $F$ represents the yield function (MPa); $a$ and $k$ represent material parameters (1/MPa and MPa, respectively) related to the rock mass cohesion $c$ and internal friction angle $\phi$, with the specific conversion relationship defined in Eq.~(\ref{eq:17}); $I_1$ represents the first stress invariant (MPa); and $J_2$ represents the second deviatoric stress invariant (MPa²). The expressions for $I_1$ and $J_2$ are:
	\begin{equation}
		\begin{cases}
			I_1=\sigma_1+\sigma_2+\sigma_3 \\
			J_2=1/6[(\sigma_1-\sigma_2)^2+(\sigma_2-\sigma_3)^2+(\sigma_3-\sigma_1)^2]
		\end{cases}
		\label{eq:5}
	\end{equation}
	where $\sigma_1$, $\sigma_2$, $\sigma_3$ are the principal effective stresses (MPa).
	
	The volumetric swelling strain increments induced by hydration and thermal effects are defined, respectively, as:
	\begin{equation}
		\begin{cases}
			d\varepsilon_{ij}^w=\delta_{ij} [k_1 dw+k_2 (dw)^2] \\
			d\varepsilon_{ij}^T=\delta_{ij} [\alpha_T dT]
		\end{cases}
		\label{eq:6}
	\end{equation}
	where $dw$ represents the water content increment (dimensionless); $dT$ represents the temperature increment (°C); $k_1$ and $k_2$ represent hydration swelling coefficients (dimensionless and 1, respectively); and $\alpha_T$ represents the thermal expansion coefficient (1/°C).
	
	To describe the degradation of mechanical properties caused by water invasion, a set of dynamic evolution relationships for rock parameters dependent on water content was introduced into the model \citep{ref25}:
	\begin{equation}
		\begin{cases}
			E=E_0 \cdot e^{-11\sqrt{(w-w_0)}} \\
			v=0.2+1.3w \\
			c=c_0-250(w-w_0) \\
			\phi=\phi_0-187.5(w-w_0)
		\end{cases}
		\label{eq:7}
	\end{equation}
	where $E$ represents the elastic modulus (MPa); $E_0$ represents the initial elastic modulus (MPa); $v$ represents the Poisson’s ratio (dimensionless); $c$ represents the cohesion (MPa); $c_0$ represents the initial cohesion (MPa); $\phi$ represents the internal friction angle (°); $\phi_0$ represents the initial internal friction angle (°); $w$ represents the current water content (dimensionless); and $w_0$ represents the initial water content (dimensionless).
	
	The spatio-temporal distribution and evolution of water content $w$ near the wellbore followed a two-dimensional diffusion equation, similar in form to the heat conduction equation \citep{ref26}:
	\begin{equation}
		\begin{cases}
			\frac{\partial w}{\partial t}=D\left(\frac{\partial^2 w}{\partial x^2}+\frac{\partial^2 w}{\partial y^2}\right) \\
			w|_{r=r_w}=w_s \\
			w|_{r\to\infty}=w_0
		\end{cases}
		\label{eq:8}
	\end{equation}
	where $D$ represents the hydration diffusion coefficient (m²/s); $w_s$ represents the saturated water content at the wellbore wall (dimensionless); $r$ represents the radial coordinate (m); and $r_w$ represents the wellbore radius (m).
	
	The pore pressure field $p$ in the formation is governed by a two-dimensional seepage equation based on Darcy’s law and the principle of mass conservation \citep{ref27}:
	\begin{equation}
		\frac{\partial}{\partial x} \left(\frac{\rho_w k}{\mu_0} \frac{\partial p}{\partial x}\right)+\frac{\partial}{\partial y} \left(\frac{\rho_w k}{\mu_0} \frac{\partial p}{\partial y}\right)+\phi\rho_w \gamma_l \frac{\partial p}{\partial t}+\rho_w \frac{\partial\phi}{\partial t}=0
		\label{eq:9}
	\end{equation}
	where $\rho_w$ represents the filtrate density (kg/m³); $k$ represents the formation permeability (m²); $\mu_0$ represents the filtrate dynamic viscosity (Pa·s); $\phi$ represents the porosity (dimensionless); $\gamma_l$ represents the liquid compressibility (1/MPa); and $t$ represents the time (s).
	
	Regarding the porosity change induced by rock skeleton deformation, the model assumes that deformation primarily affects the pore volume. The following relationship between porosity evolution and volumetric strain was established \citep{ref27}:
	\begin{equation}
		\begin{cases}
			\frac{\partial\phi}{\partial t}=\frac{(1-\phi_0)}{(1+\varepsilon_v)^2}\cdot\frac{\partial\varepsilon_v}{\partial t} \\
			\varepsilon_v=\varepsilon_x+\varepsilon_y
		\end{cases}
		\label{eq:10}
	\end{equation}
	where $\varepsilon_v$ represents the volumetric strain (dimensionless); $\phi_0$ represents the initial porosity (dimensionless); and $\varepsilon_x$ and $\varepsilon_y$ represent the principal strains (dimensionless). The coupled thermal process was constructed based on a comprehensive porous medium heat transfer equation \citep{ref28}:
	\begin{equation}
		\begin{cases}
			\lambda_{\text{eff}}=\phi\lambda_w+(1-\phi) \lambda_s \\
			\begin{multlined}[t]
				(\phi\rho_w C_M+(1-\phi) \rho_s C_s) \frac{\partial T}{\partial t} \\ -(\rho_w C_M) \frac{k}{\mu_0} \nabla p\cdot\nabla T=\nabla\cdot(\lambda_{\text{eff}}\nabla T)
			\end{multlined}
		\end{cases}
		\label{eq:11}
	\end{equation}
	where $T$ represents the temperature (°C); $C_M$ represents the specific heat capacity of the filtrate (J/(kg·K)); $\lambda_s$ represents the thermal conductivity of the rock (W/(m·K)); $\lambda_{\text{eff}}$ represents the effective thermal conductivity of the porous medium (W/(m·K)); $\lambda_w$ represents the thermal conductivity of the fluid (W/(m·K)); $\rho_s$ represents the density of the rock matrix (kg/m³); and $C_s$ represents the specific heat capacity of the rock skeleton (J/(kg·K)).
	
	\subsection{Model Assumptions and Boundary Conditions}
	\label{subsec:assumptions}
	To accurately characterize the evolution of the wellbore state under the synergistic effects of seepage, thermal, hydration, and stress multiphysics over long-term exposure, a high-fidelity numerical model was established. This model, centered on the wellbore with axisymmetric features in a polar coordinate system, was built using the COMSOL Multiphysics platform based on 2D plane-strain theory. The computational domain was selected to be much larger than the wellbore diameter to prevent far-field boundary conditions from causing non-physical interference with the near-wellbore response. In the model’s geometric setup, the wellbore was simplified to a standard circle, and the formation material was idealized as a homogeneous and isotropic porous elastic medium at the local scale. This formed the basis for applying continuum mechanics for modeling.
	
	To ensure the physical realism of the model, the application of boundary and initial conditions strictly followed the physical evolution mechanisms of the surrounding rock. For the mechanical field, a uniformly distributed radial pressure equivalent to the drilling fluid column pressure was applied to the wellbore wall. This pressure acts as the key support for preventing wellbore collapse and is a core control variable in drilling engineering. The model’s far-field boundaries were designed to replicate the in-situ stress state of the formation. This was achieved by applying loads equivalent to the maximum and minimum regional horizontal principal stresses, thereby reproducing the dominant effect of stress anisotropy on the stress concentration around the wellbore. In the seepage field, the wellbore was treated as a first-class (Dirichlet) boundary condition with a constant pressure determined by the drilling fluid column. This serves as a steady pressure source driving filtrate invasion. Conversely, the far-field boundary used a second-class (Neumann) condition of zero pore pressure gradient. This was designed to simulate an open and effectively infinite formation, ensuring that pressure disturbances propagate outward without reflection and thus avoiding boundary-induced interference. For the thermal field, considering the relatively stable temperature of circulating drilling fluid in a given section, the wellbore was also set as a constant temperature boundary, serving as a baseline for the continuous heat exchange between the wellbore and the formation. The far-field boundary maintained the initial temperature profile determined by the regional geothermal gradient, representing the undisturbed state and thus establishing the driving mechanism for heat conduction. The boundary conditions for the water content field were treated analogously to thermal diffusion. The wellbore was assigned a saturated water content to simulate the maximum hydration state, while the far-field was fixed at the initial natural water content. The resulting concentration gradient drives water absorption and diffusion, which subsequently induce the time-dependent degradation of the mechanical parameters of the rock. The initial state of the model was based on the pre-drilling formation conditions. The stress field was initialized using in-situ stress measurements, the pore pressure field was set based on hydrostatic pressure, the temperature field was constructed via linear interpolation from the geothermal gradient, and the water content field was set to the natural initial value.
	
	To focus on the core mechanisms of multiphysics coupling, several key idealized assumptions were adopted. Clarifying these assumptions is essential to properly interpret the model’s scope of applicability. For instance, in the treatment of the seepage field, the dynamic formation of a mud-cake on the wellbore wall and its significant delay effect on pressure transmission were not considered. Consequently, the simulated pressure response time is primarily determined by the intrinsic permeability of the rock mass, representing a conservative scenario without a mud-cake barrier. Regarding the physico-chemical processes, the complex rock hydration was simplified into a phenomenological model controlled by an equivalent diffusion coefficient. Additionally, the model assumes that the mechanical parameters of the rock are spatially homogeneous and isotropic, neglecting the heterogeneity and anisotropy prevalent in real sedimentary rocks. Despite these simplifications, the model effectively captures the fundamental coupling laws and response patterns of wellbore stability under the combined effects of pressure, temperature, and hydration. It also provides a high-quality dataset with intrinsic physical consistency for subsequent training of the physics-informed neural network.
	
\begin{figure}[t]
		\centering
		\includegraphics[width=\columnwidth]{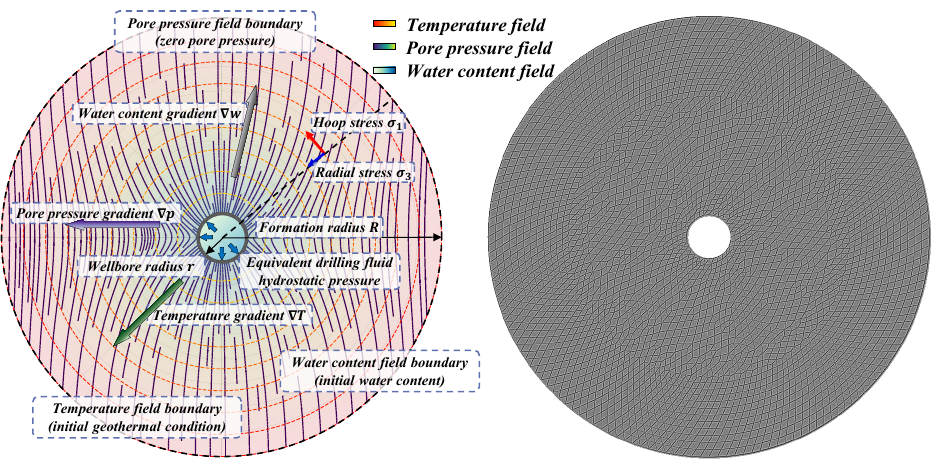}
		\vspace*{-8mm} 
		\caption{Plane Multiphysics Model and Meshing of the Wellbore.}
		\label{fig:mesh}
	\end{figure}

	During numerical discretization, an adaptive mesh refinement technique based on the physical field gradients was employed. Specifically, unstructured triangular elements were used for meshing the wellbore wall and its vicinity. This ensures sufficient resolution to capture the high-gradient physical field changes in these areas. The entire model was discretized into 6768 elements. The minimum mesh quality was 0.6292, and the average mesh quality reached 0.8898. This mesh quality is sufficient to meet the requirements for both numerical convergence and solution accuracy. The geometric configuration, the relationships between the coupled physics, and the final mesh discretization scheme of the constructed 2D multiphysics model are shown in Fig.~\ref{fig:mesh}.
	
	\subsection{COMSOL Simulation}
	\label{subsec:comsol}
	In the geosciences, constructing large-scale, high-quality datasets is recognized as a critical prerequisite for the development and evaluation of advanced artificial intelligence models \citep{ref29}. Parameter ranges were defined to systematically investigate the influence of uncertainties in formation geomechanical properties and drilling engineering parameters on the stability of the open-hole wellbore. These ranges, covering typical sandstone and mudstone lithologies, were based on field well log data and laboratory core test results from the Caofeidian block in the Bohai Sea, China. Details of the experimental characterization of the relevant mechanical parameters are provided in Appendix~A. Based on this parameter space, a high-dimensional input perturbation scheme was constructed using the Latin Hypercube Sampling (LHS) method \citep{ref30}. This approach enabled efficient and uniform sampling coverage of the entire parameter space.
	
	Given that the formation in the research block exhibits significant lithological layering, a stratified sampling strategy was further integrated within the LHS framework \citep{ref31}. This was done to accurately reflect the distinct distributions of physical and mechanical properties of the sandstone and mudstone. This sampling strategy ensures that samples are uniformly distributed within each lithological parameter subspace. It enhances the ability of the model to represent the coupled mechanisms in wellbores with multiple lithologies and improves the generalization performance of subsequent predictions. Through this method, a total of 3000 sets of input parameter samples were generated, covering over 30 key physical parameters. As shown in Table~\ref{tab:params}, the parameter ranges accounted for regional factors such as diagenetic history, degree of cementation, mineral composition, pore geometry, and temperature-pressure systems. This design ensured that the model inputs had sufficient field representativeness and practical relevance.
	
	\begin{table*}[t]
		\centering
		\caption{Sampled Ranges of Formation Parameters in the Bohai Caofeidian Block.}
		\label{tab:params}
		\begin{tabular}{lccc}
			\hline
			\textbf{Parameter Name} & \textbf{Unit} & \textbf{Sandstone} & \textbf{Mudstone} \\ \hline
			Elastic Modulus & MPa & 15640–30960 & 9600–14400 \\
			Poisson’s Ratio & - & 0.20–0.28 & 0.25–0.35 \\
			Biot’s Coefficient & - & 0.8–1.0 & 0.6–0.8 \\
			Initial Porosity & - & 0.12–0.16 & 0.07–0.11 \\
			Permeability & mD & 2.5–7.5 & 0.0005–0.002 \\
			Rock Thermal Conductivity & W/(m·K) & 1.5–2.5 & 1.0–2.0 \\
			Rock Skeleton Compressibility & 1/MPa & $(2.0–5.0)\times10^{-5}$ & $(4.0–6.0)\times10^{-5}$ \\
			Rock Skeleton Specific Heat & J/(kg·K) & 680–920 & 510–690 \\
			Fluid Compressibility & 1/MPa & $(4.0–5.0)\times10^{-4}$ & $(4.0–5.0)\times10^{-4}$ \\
			Initial Cohesion & MPa & 5–12 & 1–5 \\
			Initial Internal Friction Angle & ° & 32–38 & 22–28 \\
			Thermal Expansion Coefficient & 1/°C & $(3.0–4.0)\times10^{-5}$ & $(4.0–6.0)\times10^{-5}$ \\
			Initial Water Content & - & 0.02–0.04 & 0.02–0.04 \\
			Saturated Water Content & - & 0.08–0.16 & 0.08–0.12 \\
			Hydration Diffusion Coefficient & m²/s & $(4.0–9.0)\times10^{-10}$ & $(4.0–9.0)\times10^{-10}$ \\
			Filtrate Density & kg/m³ & 980–1020 & 980–1020 \\
			Matrix Density & kg/m³ & 2500–2700 & 2500–2700 \\
			Filtrate Viscosity & Pa·s & $(2.8–3.4)\times10^{-4}$ & $(2.8–3.4)\times10^{-4}$ \\
			Filtrate Specific Heat & J/(kg·K) & 4000–4400 & 4000–4400 \\
			Hydration Expansion Coefficient 1 & - & 0 & 0.02–0.04 \\
			Hydration Expansion Coefficient 2 & - & 0 & 0.7–0.95 \\ \hline
		\end{tabular}
	\end{table*}
	
	During the execution of the high-fidelity numerical simulations, 1910 parameter sets produced converged solutions. The output of these solutions primarily consisted of time-series distributions of the equivalent stress and plastic strain fields around the wellbore. Some parameter sets failed to converge due to numerical issues such as boundary conflicts or gradient discontinuities under strongly nonlinear coupled conditions, and these were removed from the dataset. The resulting evolution sequences fully capture the stability evolution patterns under typical wellbore exposure conditions. This provides strong, physics-constrained sample support for subsequent training of the physics-informed neural network. It also offers a data foundation for optimizing drilling fluid parameters to enhance wellbore stability.
	
	\begin{figure*}[t]
		\centering
		\includegraphics[width=\textwidth]{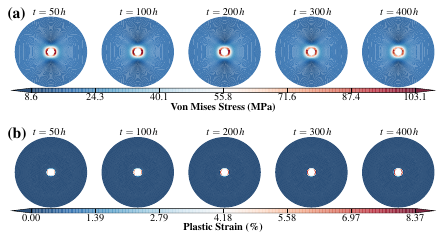}
		\vspace*{-10mm} 
		\caption{Evolution of Von Mises Stress Field and Plastic Strain Field at Different Times.}
		\label{fig:vonmises_strain}
	\end{figure*}
	
	Fig.~\ref{fig:vonmises_strain} shows the 2D distribution and evolution of the Von Mises equivalent stress field and the plastic strain field in the rock surrounding the open-hole wellbore at different evolution times. As shown in Fig.~\ref{fig:vonmises_strain}(a), the Von Mises equivalent stress exhibits a nearly symmetric distribution in the near-wellbore region during the initial stage. The high-stress concentration is mainly confined to the area immediately adjacent to the wellbore, with a maximum stress value of approximately 80~MPa. As drilling fluid continuously invades, causing an increase in pore pressure, combined with heat conduction and hydration swelling, a rapid local stress accumulation occurs in the wellbore region by $t = 100$~h. This forms a distinct annular high-stress zone. The maximum stress increases to 102.96~MPa, reaching the peak value for the entire process. This represents a state of rapid stress concentration and reduced shear strength in the near-wellbore region due to the combined effects of temperature rise, pore pressure increase, and hydration swelling. As the evolution proceeds to $t = 200–400$~h, the hydration process continues, and pore pressure gradually diffuses and equilibrates. Stress is redistributed and released in local areas. The peak Von Mises stress slightly decreases, and the overall distribution tends to stabilize, indicating that the surrounding rock has entered a phase of stress relief and redistribution.
	
	Fig.~\ref{fig:vonmises_strain}(b) shows the spatio-temporal evolution of the plastic strain field in the surrounding rock. It exhibited a synergistic evolution with the Von Mises stress field but with different temporal characteristics. In the early stage, the plastic strain value was close to zero, with only a weak response in the near-wellbore region. By $t = 100$~h, as local areas began to yield, the plastic strain rapidly grew to 6.90\%, forming a localized high-strain zone. This reflects the initial onset and local accumulation of plastic failure. Subsequently, within the period of $t = 200–400$~h, the ongoing hydration process caused the plastic zone to expand further. The plastic strain value continued to increase, reaching a maximum of 8.37\%, and formed a stable, closed plastic annulus structure. This process indicates that plastic failure continuously expands under high-pressure, high-temperature, and hydration conditions, while stress at the wellbore wall gradually decreases toward a stable state.
	
	\subsection{Analysis of Simulation Results and Physical Consistency Check}
	\label{subsec:validation}
\begin{figure*}[t]
		\centering
		\includegraphics[width=\textwidth]{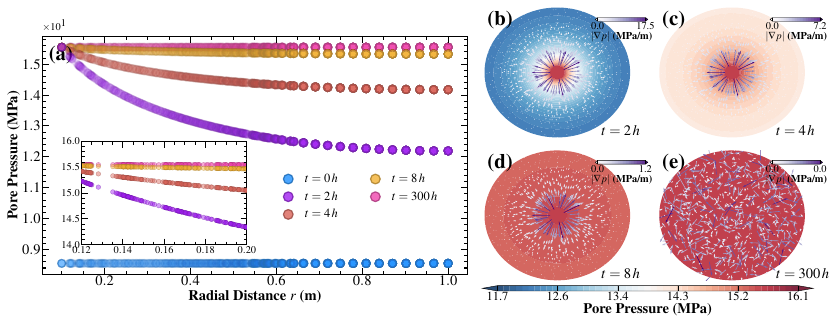}
		\vspace*{-10mm} 
		\caption{Evolution of Pore Pressure Radially and the Pore Pressure Field at Different Times.}
		\label{fig:pore_pressure}
	\end{figure*}

\begin{figure*}[t]
		\centering
		\includegraphics[width=\textwidth]{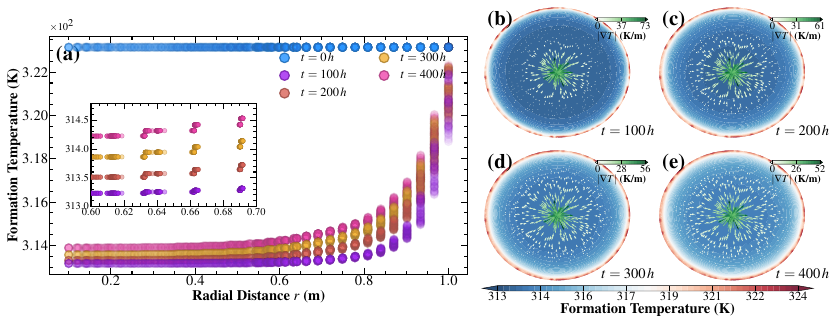}
		\vspace*{-10mm} 
		\caption{Evolution of Temperature Radially and the Temperature Field at Different Times.}
		\label{fig:temperature}
	\end{figure*}

\begin{figure*}[t]
		\centering
		\includegraphics[width=\textwidth]{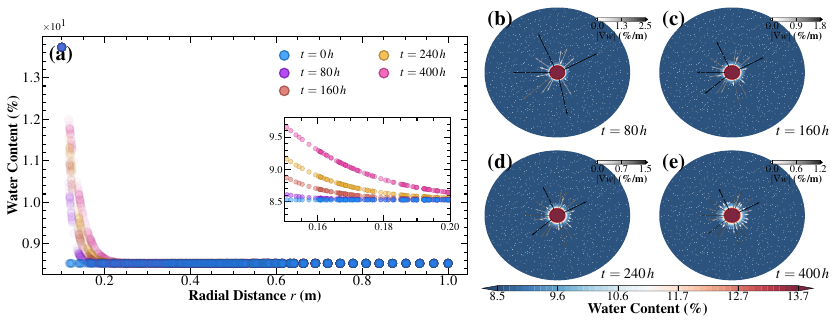}
		\vspace*{-10mm} 
		\caption{Evolution of Water Content Radially and the Water Content Field at Different Times.}
		\label{fig:water_content_space}
	\end{figure*}

	To evaluate the physical plausibility and numerical accuracy of the constructed Multiphysics-coupled model in characterizing the response of the rock around the wellbore, Figs.~\ref{fig:pore_pressure} to \ref{fig:water_content_space} show the radial and spatial evolution of pore pressure, temperature, and water content at different times under continuous drilling fluid invasion.
	
	Fig.~\ref{fig:pore_pressure}(a) shows the evolution of pore pressure along the radial direction at different times. The results revealed a diffusion-dominated seepage evolution mechanism in the open-hole section under continuous drilling fluid invasion. At $t = 0$~h, the pore pressure in the near-wellbore region was significantly below the high-pressure threshold and was confined to the immediate vicinity of the wellbore, indicating that the mud support had not yet formed an effective pressure drop. By $t = 8$~h, the high-pressure zone rapidly expanded to cover the entire radial domain. The pressure distribution gradually became smoother, and the gradient weakened. After $t = 300$~h, the pressures near the wellbore and in the far-field were nearly identical. The seepage disturbance had completed its full-domain diffusion and reached a stable state. The 2D pore pressure field evolutions in Fig.~\ref{fig:pore_pressure}(b) to 4(e) further confirm this radial diffusion trend. They show concentric, high-pressure contours gradually expanding outward. In the early stage, the high-pressure zone was limited to the wellbore vicinity. As evolution progressed, this high-pressure area became enlarged, and the zone of dense contours expanded significantly. In the late stage, the density of the pressure contours decreased, and the pressure field approached a uniform distribution. Minor local fluctuations in the gradient reflected the effect of the in-situ stress orientation on the seepage disturbance. This behavior aligns with the predictions of Biot-Darcy coupling theory, verifying both the physical consistency and the accuracy of the model in capturing the seepage-mechanics response process.
	
	Fig.~\ref{fig:temperature}(a) shows the evolution of temperature along the radial direction at different times. Under the continuous cooling effect of the drilling fluid, the open-hole section exhibited non-steady-state heat transfer characteristics dominated by thermal diffusion. The near-wellbore region initially recorded a lower temperature due to the cooling from the drilling fluid, while the far-field formation acted as a high-temperature heat source. Heat was continuously conducted from this high-temperature far-field region toward the low-temperature near-wellbore region. This process caused the temperature in the near-wellbore area to gradually recover over time, and the temperature difference between it and the far-field progressively decreased. The 2D temperature fields in Fig.~\ref{fig:temperature}(b) to 5(e) further confirmed this trend. The temperature fields showed that heat from the high-temperature far-field was conducted along the temperature gradient toward the wellbore. The low-temperature zone gradually warmed up, and the temperature difference diminished. The isotherms advanced stably in a concentric circular pattern. The direction of the temperature gradient was consistent with the direction of the principal in-situ stress. The heat transfer process as a whole exhibited nearly axisymmetric diffusion characteristics, verifying the physical consistency and accuracy of the model in simulating the thermo-fluid coupling process in the formation.
	
	Fig.~\ref{fig:water_content_space}(a) shows the evolution of water content along the radial direction at different times. It reflects the slow accumulation of water in the rock surrounding the open hole, a process controlled by hydration-diffusion. In the initial stage, the water content near the wellbore was only slightly higher than the far-field background value. As time progressed, the water content near the wellbore slowly increased. However, the high-water-content zone remained confined to the near-wellbore region and did not diffuse extensively into the far-field. By $t = 400$~h, the high-water-content zone had not significantly expanded beyond the immediate vicinity of the wellbore. The water content distribution exhibited typical slow seepage-diffusion characteristics. Furthermore, the water content in the far-field did not continuously rise to approach the wellbore value. Instead, through a process of absorption, diffusion, and redistribution, it gradually trended toward its own stable formation water content level. This demonstrates a limited and restricted range of hydration influence. The 2D water content field evolutions in Fig.~\ref{fig:water_content_space}(b) to 6(e) further confirmed the above analysis from a spatial perspective. The water content field consistently showed a localized high-value region centered at the wellbore. The extent of hydration-diffusion was limited, and its delayed effect was significant, verifying the physical consistency and accuracy of the model in the coupled simulation of hydration-diffusion and local mechanical softening of the surrounding rock.
	
	\begin{figure}[t]
		\centering
		\includegraphics[width=\columnwidth]{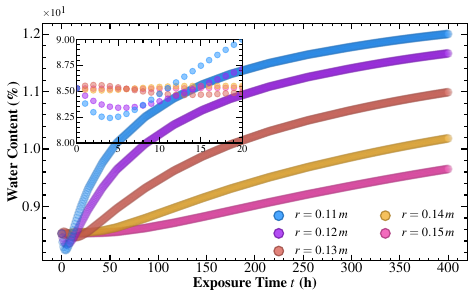}
		\vspace*{-9mm} 
		\caption{Relationship between Water Content and Time at Different Radial Locations.}
		\label{fig:water_content_time}
	\end{figure}
	
	Fig.~\ref{fig:water_content_time} shows the dynamic change of water content over time at different radial locations. It indicates that the hydration-diffusion front steadily advances outward from the near-wellbore region during the evolution. The increase in water content is most significant in the near-wellbore region, while the water absorption trend in the far-field is slow and delayed. The overall process exhibits a typical effect of spatial gradient control. In the initial stage ($t < 4$~h), there were slight fluctuations in water content at some locations. This was mainly attributed to the transient adjustment effects of initial field disturbances or numerical relaxation mechanisms in the early period. As time increased, the system quickly stabilized and entered a phase of monotonic accumulation. Overall, the dynamic evolution of water content along the radial direction demonstrates diffusion-controlled accumulation and temporal response characteristics, confirming the physical consistency and accuracy of the model in simulating the hydration-diffusion process.
	
	\section{Development of the Multiphysics Physics-Informed Neural Network}
	\label{sec:pinn_development}
	
	\subsection{Model Architecture Design}
	\label{subsec:architecture}
	
	\subsubsection{Network Architecture Design}
	A novel network architecture, termed STWM-PINN, was designed and constructed to efficiently simulate the evolution of wellbore states under synergistic “seepage-thermal-hydration-mechanics” multiphysics disturbances. The model also enables high-precision prediction of key state variables. This architecture is based on Physics-Informed Neural Networks (PINNs), a deep learning framework that integrates physical laws with a data-driven learning paradigm \citep{ref32}. The core mechanism of PINNs involves integrating the residuals of partial differential equations (PDEs), along with corresponding boundary conditions (BCs) and initial conditions (ICs), into the network’s loss function. This design allows the network to be trained effectively using only physical prior information in a “weakly-supervised” manner, even when labeled data are extremely sparse or entirely missing. This characteristic gives PINNs a significant advantage over traditional data-driven neural networks when dealing with complex geo-engineering problems where data acquisition is expensive and incomplete.
	
	The fundamental principle of the PINN method is to construct an approximate solution to a multiphysics problem using a deep neural network, $u_{\theta}(x,t)$, characterized by a set of learnable parameters $\theta$. The network receives spatio-temporal coordinates $\mathbf{x}=(x,y,t)$ as input and outputs a vector of state variables $\mathbf{u}=(p,T,w,\sigma_{\text{eq}},\varepsilon_p)$ corresponding to that point. Using automatic differentiation, any order of derivatives of the output $\mathbf{u}$ with respect to the inputs $\mathbf{x}$ and $t$ can be calculated efficiently. This allows the residual of the governing physical equations to be formulated as a core loss term to guide the network’s training process.
	
	Regarding specific implementation, PINN models commonly employ a Fully Connected Feedforward Neural Network (FNN) \citep{ref33}. The forward propagation of information in this network follows the recursive relationship below:
	\begin{equation}
		\begin{cases}
			\mathbf{a}^l = \mathbf{W}^l \mathbf{z}^{(l-1)} + \mathbf{b}^l \\
			\mathbf{z}^l = \sigma(\mathbf{a}^l) \\
			\mathbf{f}_{NN} = \mathbf{W}^{(L+1)} \mathbf{z}^L + \mathbf{b}^{(L+1)}
		\end{cases}
		\label{eq:12}
	\end{equation}
	where $\mathbf{W}^l$ and $\mathbf{b}^l$ represent the weight matrix and bias vector of the $l$-th layer, respectively; $\sigma(\cdot)$ represents the activation function; $L$ represents the total number of hidden layers; and $\mathbf{f}_{NN}$ represents the final output. Through optimization algorithms such as backpropagation and gradient descent, the network parameters are iteratively updated. This drives the network’s output to approximate available observational data and strictly adhere to the governing physical equations and associated constraints.
	
	To ensure that the network output simultaneously satisfies physical laws and well-posed conditions, the total loss function of a PINN is generally composed of multiple components:
	\begin{equation}
		L_{\text{total}}=\omega_f L_{\text{PDE}}+\omega_b L_{\text{BC}}+\omega_0 L_{\text{IC}}+\omega_d L_{\text{Data}}
		\label{eq:13}
	\end{equation}
	where $L_{\text{PDE}}$ represents the mean squared error of the governing equation residuals at a selection of collocation points in the domain; $L_{\text{BC}}$ and $L_{\text{IC}}$ correspond to the residuals of the boundary and initial conditions, respectively; and $L_{\text{Data}}$ represents the fitting error between the model’s predictions and available labeled data points. Each loss component is scaled by a corresponding weight coefficient $\omega$ to balance its relative contribution to the network’s training and optimization process.
	
	Automatic Differentiation (AD) is essential for computing the $L_{\text{PDE}}$ residual term \citep{ref34}. It allows the partial derivatives of the network’s output with respect to its input variables to be obtained accurately and efficiently through the chain rule via the backpropagation mechanism. This avoids the need for explicit construction of finite difference schemes or symbolic derivative expressions. This advantage significantly improves both training efficiency and differentiation accuracy. It makes PINNs particularly effective for handling multiphysics problems involving high-order derivative terms or strong nonlinearities, and particularly suitable for solving complex PDE systems and inverting their spatio-temporal dynamic solutions.
	
	Regarding the training mechanism, PINNs employ a “weakly-supervised” strategy based on collocation point sampling. The satisfaction of the governing equations is treated as a training objective, with the residual penalty distributed over the entire spatio-temporal domain. This approach bypasses the dependence on structured meshes required by traditional numerical methods, thus avoiding associated numerical stability issues and memory burdens. This allows PINNs to perform well in unstructured domains, with complex boundary conditions, and for problems involving nonlinear degradation. Furthermore, the training framework offers the flexibility to incorporate sparse observational data or high-fidelity simulation results through the $L_{\text{Data}}$ term. This strengthens the model’s ability to fit specific local regions, boundary behaviors, or key time points, thereby enhancing its adaptability to realistic geological conditions.
	
	Building on this foundation, the proposed STWM-PINN model extends the standard PINN framework by introducing a multi-branch physics prediction architecture. Relatively independent neural network sub-modules were designed for each of the four core physical processes: seepage, heat transfer, hydration diffusion, and elasto-plastic mechanics. These sub-modules are ultimately trained jointly through a unified loss function composed of the governing equation residuals, which structurally ensures the consistency of the coupling relationships between the different physical fields.
	
	\textbf{Seepage Prediction Branch:} This sub-network is dedicated to solving the governing equation for the pore pressure $p$ distribution. Its training is primarily driven by the PDE residual to achieve learning guided purely by physical laws. Additionally, to enhance the fitting accuracy in specific critical regions, a small amount of pressure data generated by COMSOL Multiphysics was introduced as a data supervision term to assist the training.
	
	\textbf{Thermal Prediction Branch:} The training of this branch is based on the heat transfer equation as its core constraint. It primarily utilizes the thermal diffusion residual, generated by applying automatic differentiation to the predicted temperature $T$ with respect to the spatio-temporal coordinates, to simulate the complex non-steady-state heat exchange between the wellbore wall and the formation.
	
	\textbf{Hydration Prediction Branch:} This branch is responsible for predicting the spatio-temporal evolution of water content $w$. Its training constraint is a diffusion loss term constructed by analogy to the heat transfer equation. A low diffusion coefficient is specifically imposed as a constraint to enhance the model’s ability to capture the slow, delayed nature of the hydration process in the near-wellbore region. To prevent information leakage from the high-fidelity data, the hydration and thermal prediction branches are trained without using gradient information from the simulation data. Their physics-based learning path is constructed solely from the initial/boundary conditions and the equation residuals.
	
	\textbf{Mechanical Output Branch:} This module internally couples the Drucker–Prager yield criterion with the elasto-plastic constitutive relationship. It uses the water content $w$ and temperature $T$ predicted by the other branches to dynamically update the rock’s mechanical parameters. It ultimately outputs the equivalent stress $\sigma_{\text{eq}}$ and plastic strain $\varepsilon_p$. The output of this module is also constrained by the corresponding mechanical governing equation residuals, and its predictions can be directly compared with the COMSOL Multiphysics simulation data to evaluate its final predictive accuracy.
	
\begin{figure*}[t]
		\centering
		\includegraphics[width=0.9\textwidth]{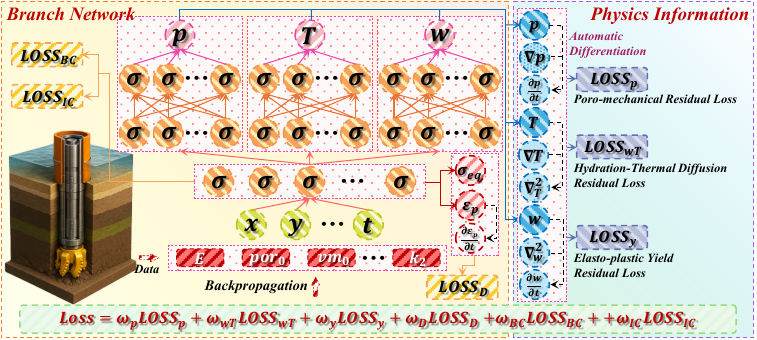}
		\vspace*{-3mm} 
		\caption{Architecture of the STWM-PINN Algorithm.}
		\label{fig:pinn_arch}
	\end{figure*}

	This multi-branch, synergistic prediction architecture enables the STWM-PINN to effectively model multiphysics problems while ensuring that the coupling between the different physical processes is satisfied within the network. The overall network architecture is illustrated in Fig.~\ref{fig:pinn_arch}.
	
	\subsubsection{Embedding of Physics Constraints}
	To ensure that the neural network’s predictions maintain physical consistency and mechanistic interpretability when simulating the multiphysics response of the open-hole wellbore, the STWM-PINN embeds the residuals of the multiphysics governing equations as core penalty terms within its training framework. All residuals are computed using AD, which eliminates the need for finite difference schemes and ensures numerical stability and physical accuracy for high-order derivative calculations. The construction of the residual loss terms corresponds to the previously established seepage-thermal-hydration-stress coupled governing equation system. The primary components include the seepage-mechanics coupled residual, the hydration-thermal diffusion coupled residual, the elasto-plastic failure residual, and the boundary and initial condition error terms.
	
	For the seepage and mechanics coupling, the following physics constraint was constructed based on Darcy’s law and the principle of mass conservation:
	\begin{equation}
		\begin{split}
			L_p = \left\lVert \right. & \frac{\partial}{\partial x} \left(\frac{\rho_w k}{\mu_0} \frac{\partial p}{\partial x}\right) + \frac{\partial}{\partial y} \left(\frac{\rho_w k}{\mu_0} \frac{\partial p}{\partial y}\right) \\
			& \left. \quad + \phi\rho_w \gamma_l \frac{\partial p}{\partial t}+\rho_w \frac{\partial\phi}{\partial t}\right\rVert_2^2
		\end{split}
		\label{eq:14}
	\end{equation}
	where $L_p$ represents the seepage-mechanics residual loss (dimensionless). The residual characterizes spatial diffusion, temporal evolution, and porosity-driven changes in the pore pressure field within a 2D porous medium. During training, the neural network uses AD to compute the various derivatives of pore pressure $p$ with respect to space and time. It is also coupled with the volumetric-strain-based porosity evolution function (see Eq.~(\ref{eq:10})) to ensure the dynamic consistency between the pore pressure and rock skeleton deformation responses.
	
	For the coupled hydration and thermal diffusion processes, the loss function consists of two parts. These correspond to the hydration model, which is analogous in form to heat conduction, and the non-steady-state heat transfer governing equation. The joint expression is:
	\begin{equation}
		\begin{split}
			L_{wT} = & \left\lVert\frac{\partial w}{\partial t}-D\left(\frac{\partial^2 w}{\partial x^2}+\frac{\partial^2 w}{\partial y^2}\right)\right\rVert_2^2 \\
			& + \left\lVert (\phi\rho_w C_M+(1-\phi) \rho_s C_s) \frac{\partial T}{\partial t} \right. \\
			& \qquad \left. - (\rho_w C_M) \frac{k}{\mu_0} \nabla p\cdot\nabla T-\nabla\cdot(\lambda_{\text{eff}}\nabla T)\right\rVert_2^2
		\end{split}
		\label{eq:15}
	\end{equation}
	where $L_{wT}$ represents the hydration-thermal diffusion residual loss (dimensionless). The first term characterizes the spatiotemporal diffusion of water content $w$ within the rock mass, driven by drilling fluid invasion. The second part strictly enforces the complete energy conservation law, which includes the combined heat capacity of the rock-fluid system, convective heat transfer by the fluid, and effective thermal conduction through the porous medium. To improve the ability of the model to distinguish between different physical processes and to prevent information leakage, gradient information from the COMSOL Multiphysics simulations for the thermal and hydration fields was intentionally withheld during the training phase. This forces the network to learn the physics solely from the boundary/initial conditions and the equation residuals.
	
	For the elasto-plastic mechanical behavior, the constraint was constructed using the classic form of the Drucker–Prager yield criterion as its physics residual:
	\begin{equation}
		L_y=\lVert aI_1+\sqrt{J_2}-k \rVert_2^2
		\label{eq:16}
	\end{equation}
	where $L_y$ represents the elasto-plastic yield residual loss (dimensionless); $I_1$ represents the first principal stress invariant (MPa), whose definition and calculation method are inherited from Eq.~(\ref{eq:5}); $J_2$ is the second deviatoric stress invariant (MPa²), inherited from Eq.~(\ref{eq:5}); and $a$ and $k$ represent the Drucker-Prager yield surface parameters, defined as \citep{ref35}:
	\begin{equation}
		\begin{cases}
			a = \frac{2 \sin\phi(w)}{\sqrt{3} (3-\sin\phi(w))} \\
			k = \frac{6c(w) \cos\phi(w)}{\sqrt{3} (3-\sin\phi(w))}
		\end{cases}
		\label{eq:17}
	\end{equation}
	where $c(w)$ represents the cohesion function controlled by water content (MPa), and $\phi(w)$ represents the internal friction angle function controlled by water content (°). The values of cohesion $c(w)$ and internal friction angle $\phi(w)$ are dynamically adjusted according to the functional relationship defined in Eq.~(\ref{eq:7}). This ensures that the mechanics module can sensitively respond to the evolution of rock strength under the influence of hydration and thermal effects.
	
	In addition to the physics-based residuals from the governing equations, the boundary condition error $L_{BC}$ and initial condition error $L_{IC}$ were also incorporated into the total loss function as explicit constraints. This is intended to drive the model’s output to match the specified boundary and initial states. The boundary condition error is defined as:
	\begin{equation}
		\begin{split}
			L_{BC} = \sum_{i=1}^{N_{BC}} \sum_{j=1}^{N_t} \Biggl[ & (p_{\text{pred}}(x_i^b,t_j)-p_{BC}(x_i^b,t_j))^2 \\
			& + (T_{\text{pred}}(x_i^b,t_j)-T_{BC}(x_i^b,t_j))^2 \\
			& + (w_{\text{pred}}(x_i^b,t_j)-w_{BC}(x_i^b,t_j))^2 \Biggr]
		\end{split}
		\label{eq:18}
	\end{equation}
	where $p_{\text{pred}}$, $T_{\text{pred}}$, and $w_{\text{pred}}$ represent the network’s predicted pore pressure, temperature, and water content (units of MPa, °C, and dimensionless, respectively); $p_{BC}$, $T_{BC}$, and $w_{BC}$ represent the ground truth values at the boundary conditions, with the same units as the predicted values; $N_{BC}$ represents the total number of boundary sampling points; $N_t$ represents the total number of discrete time sampling points; $x_i^b$ represent the spatial sampling points on the model’s boundary; and $t_j$ represent the discrete time points (s).
	
	The initial condition error is defined similarly:
	\begin{equation}
		\begin{split}
			L_{IC} = \sum_{i=1}^{N_{IC}} \Biggl[ & (p_{\text{pred}}(x_i^0,t_0)-p_{IC}(x_i^0,t_0))^2 \\
			& + (T_{\text{pred}}(x_i^0,t_0)-T_{IC}(x_i^0,t_0))^2 \\
			& + (w_{\text{pred}}(x_i^0,t_0)-w_{IC}(x_i^0,t_0))^2 \Biggr]
		\end{split}
		\label{eq:19}
	\end{equation}
	where $p_{IC}$, $T_{IC}$, and $w_{IC}$ represent the ground truth values at the initial time; $N_{IC}$ represents the total number of initial condition sampling points; $t_0$ represents the initial time (s); and $x_i^0$ represent the spatial sampling points at the initial time.
	
	\subsection{Model Training}
	\label{subsec:training}
	To ensure the STWM-PINN model achieves physical consistency, numerical stability, and good convergence efficiency in its prediction task, the overall training process aims to optimize a composite loss function. The construction of this loss function uses a weighted mean squared error structure. Its components include the residuals of the multiphysics governing equations, the errors from the boundary and initial condition constraints, and the fitting error from the supervised learning data labels. The specific expression for this composite loss function is as follows:
	\begin{equation}
		\begin{split}
			L_{\text{total}} ={} & \omega_p \cdot L_p + \omega_{wT} \cdot L_{wT} + \omega_y \cdot L_y \\
			& + \omega_{\text{data}} \cdot L_{\text{data}} + \omega_{BC} \cdot L_{BC} + \omega_{IC} \cdot L_{IC}
		\end{split}
		\label{eq:20}
	\end{equation}
	where $L_p$ represents the seepage-mechanics residual loss; $L_{wT}$ represents the hydration-thermal diffusion residual loss; $L_y$ represents the elasto-plastic yield residual loss; $L_{\text{data}}$ represents the supervised data label error, and $L_{BC}$ and $L_{IC}$ represent the physical residual terms for the boundary and initial conditions, respectively.
	
	To balance the contributions of the different physics constraints during the network optimization process, the weight coefficients $\omega$ for each loss term, it must be carefully set. These weights were determined empirically based on a comprehensive consideration of the numerical magnitude of the different physics residuals, their gradient sensitivity to the parameters, and their importance to the overall task. The specific values were set as follows:
	\begin{equation}
		\begin{cases}
			\omega_p = 1.2 \times 10^{-4} \\
			\omega_{wT} = 1.1 \times 10^{-12} \\
			\omega_y = 2.0 \times 10^{-4} \\
			\omega_{\text{data}} = \omega_{BC} = \omega_{IC} = 1.0
		\end{cases}
		\label{eq:21}
	\end{equation}
	For the choice of optimization algorithm, the AdamW optimizer was adopted \citep{ref36}. To prevent model overfitting and enhance its generalization performance, two regularization techniques, Weight Decay and Dropout, were simultaneously integrated into the training \citep{ref37}. The GELU activation function was selected, as its smooth characteristics are beneficial for promoting the synergistic convergence of the multiphysics residual terms \citep{ref38}.
	
	To guide the model training to effectively escape local minima and approach a better solution, a dynamic learning rate adjustment strategy was employed, namely, three-stage Cosine Annealing with Restarts \citep{ref39}. At any given step $t$, the learning rate $\eta_t$ is given by:
	\begin{equation}
		\eta_t = \eta_{\min} + \frac{1}{2} (\eta_{\max}^{(s)} - \eta_{\min}) \left[1+\cos\left(\frac{\pi \cdot t_{\text{cur}}}{T^{(s)}}\right)\right]
		\label{eq:22}
	\end{equation}
	where $\eta_{\max}^{(s)}$ represents the maximum learning rate for the $s$-th stage; $\eta_{\min}$ represents the minimum learning rate during training, and $t_{\text{cur}}$ and $T^{(s)}$ represent the number of steps already taken in the current stage and the total number of steps in that stage, respectively. In addition, to avoid overfitting, the training process was also monitored by an Early Stopping strategy, which determines whether to end the training prematurely by tracking the changes in the loss on a validation set.
	
	The core hyperparameter configurations used in the training of the STWM-PINN model are summarized in Table~\ref{tab:hyperparams}.
	
	\begin{table*}[t]
		\centering
		\caption{Hyperparameter Configuration for STWM-PINN Model Training.}
		\label{tab:hyperparams}
		\begin{tabular}{llc}
			\hline
			\textbf{Symbol} & \textbf{Description} & \textbf{Value} \\ \hline
			$N_{BC}$ & Number of boundary condition sampling points & 800 \\
			$N_{IC}$ & Number of initial condition sampling points & 6768 \\
			$N_{R}$ & Number of physics residual sampling points (each) & 6768 \\
			$N_{\text{data}}$ & Number of supervised data sampling points & 3000 \\
			$B$ & Batch size & 256 \\
			$L$ & Number of hidden layers & 6 \\
			$N_h$ & Neurons per layer & 128 \\
			$\sigma(\cdot)$ & Activation function & GELU \\
			$\omega_p, \omega_{wT}, \omega_y$ & Multiphysics residual term weights & $[1.2\times10^{-4}, 1.1\times10^{-12}, 2.0\times10^{-4}]$ \\
			$\omega_{\text{data}}, \omega_{BC}, \omega_{IC}$ & Data and boundary term weights & $[1.0, 1.0, 1.0]$ \\
			$[T_1,T_2,T_3]$ & Steps for three-stage cosine annealing & [25k, 45k, 30k] \\
			$\eta_{\text{stage\_max}}$ & Initial learning rate for each stage & $[1.0\times10^{-3}, 5.0\times10^{-3}, 2.0\times10^{-4}]$ \\
			$\eta_{\min}$ & Minimum learning rate & $1.00\times10^{-5}$ \\
			$\lambda_{wd}$ & Weight decay coefficient & $1.00\times10^{-4}$ \\
			$p_d$ & Dropout rate & 0.2 \\
			$N_{\text{patience}}$ & Early Stopping patience rounds & 5000 \\
			$\delta_{\text{stop}}$ & Early Stopping threshold & $1.00\times10^{-6}$ \\ \hline
		\end{tabular}
	\end{table*}
	
	\begin{table*}[t]
		\caption{STWM-PINN Algorithm Flow.}
		\label{tab:algorithm}
		\begin{algorithmic}[1]
			\State \textbf{Input:}
			\State $X_R$: Set of physics residual points (total $N_R$)
			\State $X_Y, Y_{\text{label}}$: Set of supervised data points (total $N_{\text{data}}$)
			\State $X_{IC}$: Set of initial condition points (total $N_{IC}$)
			\State $X_{\text{val}}$: Validation set for early stopping
			\State $\Theta_{NN}^{(0)}$: Initial parameters of the neural network
			\State $H$: Set of hyperparameters (e.g., $L, N_h, B, \omega, \eta, ...$)
			\State \textbf{Output:}
			\State $\Theta_{NN}$: Trained network parameters after convergence
			\State $N(t,x,y;\Theta_{NN})$: The corresponding state variable predictor
			\Function{STWM-PINN}{}
			\State \textbf{Initialize:} Initialize the neural network $N(t,x,y;\Theta_{NN})$ with architecture defined by hyperparameters $L, N_h, \sigma(\cdot), p_d$
			\State \textbf{Construct the composite loss function:} $L_{\text{total}} =\omega_p L_p+\omega_{wT} L_{wT}+\omega_y L_y+\omega_{\text{data}} L_{\text{data}}+\omega_{BC} L_{BC}+\omega_{IC} L_{IC}$
			\State \textbf{Set the weights for each loss term:} $\omega=\{\omega_p,\omega_{wT},\omega_y,\omega_{\text{data}},\omega_{BC},\omega_{IC}\}$
			\State \textbf{Initialize} the AdamW optimizer and set the weight decay $\lambda_{wd}$
			\State \textbf{Initialize} the learning rate scheduler with parameters ($T_s, \eta_{\text{stage\_max}}, \eta_{\min}$)
			\For{epoch = 1 to $10^5$}
			\State Sample a training batch of size $B$ from the total point sets $X_R, X_Y, X_{BC}, X_{IC}$
			\State Compute the loss terms on the batch data via forward propagation and AD: $L_p, L_{wT}, L_y, L_{\text{data}}, L_{BC}, L_{IC}$
			\State Calculate the total loss $L_{\text{total}}$ by a weighted sum according to $\omega$
			\State Update the current learning rate $\eta_t$ according to the scheduler
			\State Update network parameters using the AdamW optimizer: $\Theta_{NN} \leftarrow \Theta_{NN} - \eta_t \cdot \nabla_{\Theta_{NN}} L_{\text{total}}$
			\State Periodically compute the validation loss $L_{\text{val}}$ on the validation set $X_{\text{val}}$
			\If{$L_{\text{val}}$ has not improved by less than $\delta_{\text{stop}}$ for $N_{\text{patience}}$ consecutive steps}
			\State \textbf{break} \Comment{Trigger early stopping and terminate training}
			\EndIf
			\EndFor
			\State Use the final trained parameters $\Theta_{NN}$ to generate the state variable predictor
			\State \Return $\Theta_{NN}, N(t,x,y;\Theta_{NN})$
			\EndFunction
		\end{algorithmic}
	\end{table*}
	
	Table~\ref{tab:algorithm} shows the execution logic of the STWM-PINN algorithm for state prediction, physics constraint embedding, and training iteration.
	
\begin{figure*}[t]
		\centering
		\includegraphics[width=\textwidth]{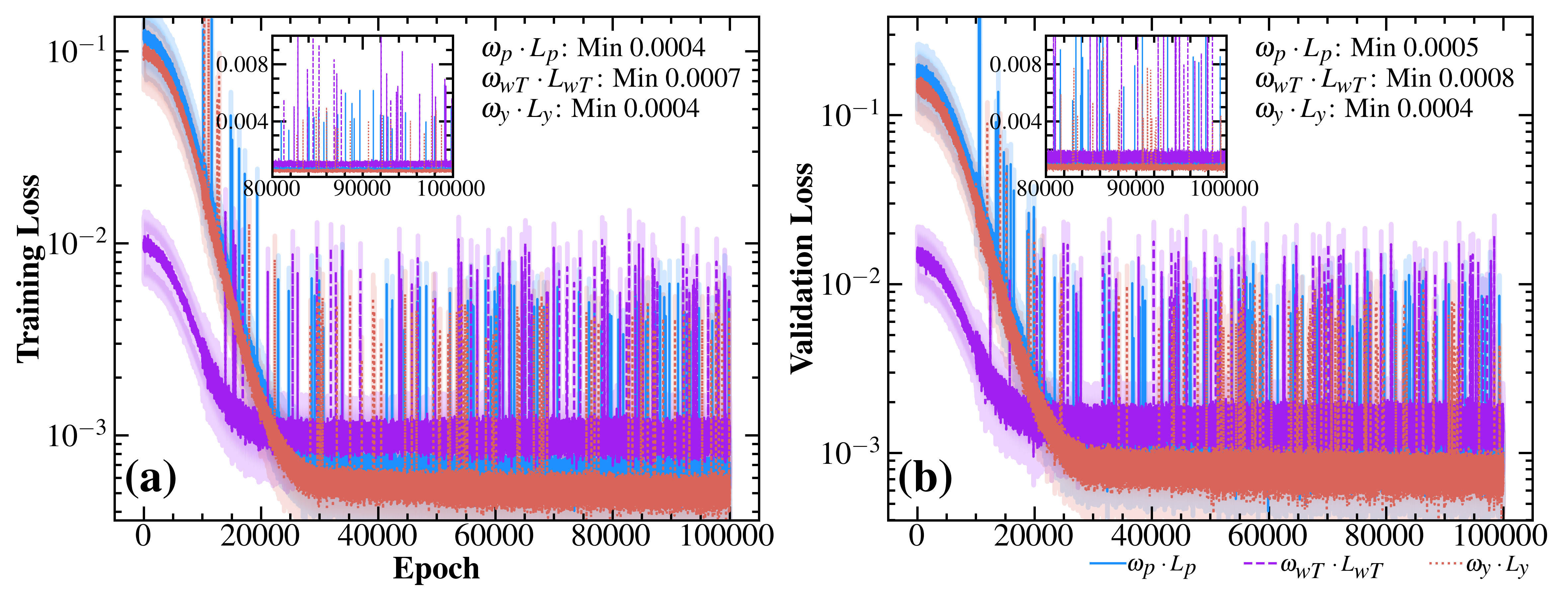}
		\vspace*{-10mm} 
		\caption{Training and Validation Loss Curves for the Main Physics Residual Terms of the STWM-PINN Model.}
		\label{fig:loss_residuals}
	\end{figure*}

\begin{figure*}[t]
		\centering
		\includegraphics[width=\textwidth]{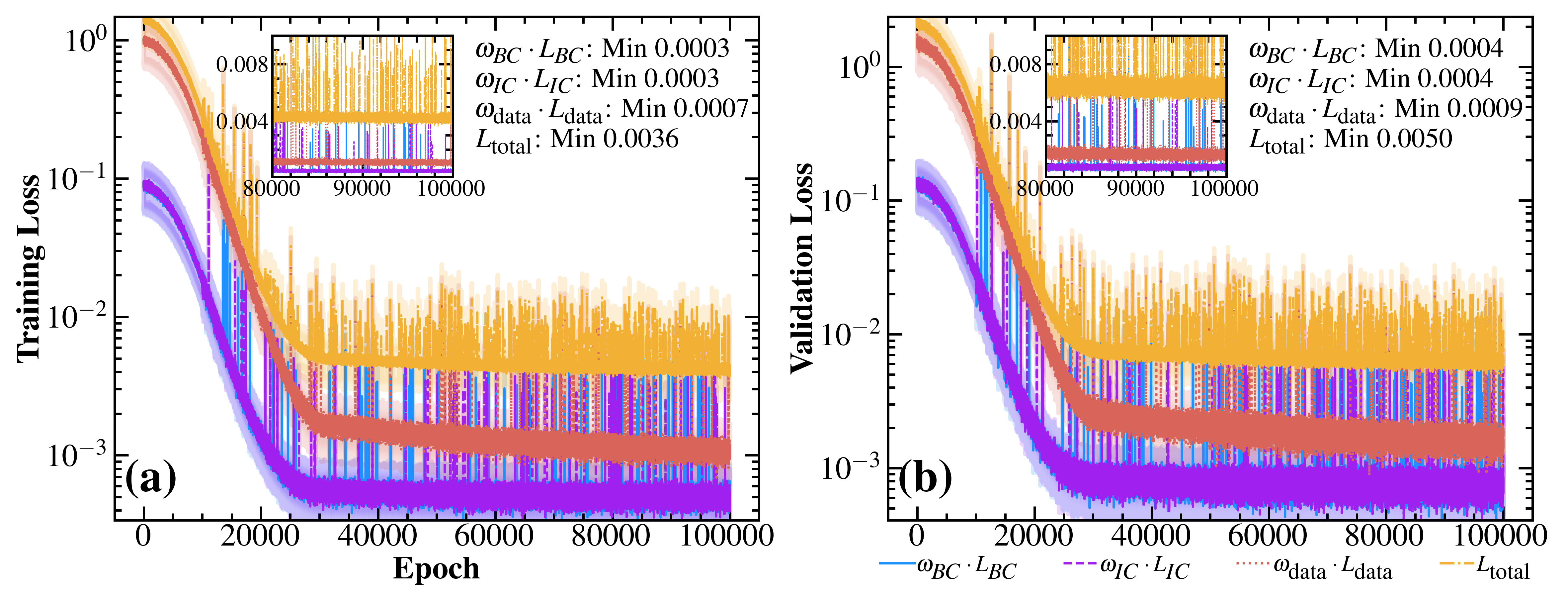}
		\vspace*{-10mm} 
		\caption{Training and Validation Loss Curves for the Boundary Condition, Initial Condition, and Data Fitting Terms of the STWM-PINN Model.}
		\label{fig:loss_bcs}
	\end{figure*}

	To evaluate the physics-consistent fitting performance and the generalization capability of the STWM-PINN on unseen samples during the training process, Fig.~\ref{fig:loss_residuals} and Fig.~\ref{fig:loss_bcs} show the evolution trends of the various physics residual terms, the boundary condition loss, the initial condition loss, and the total loss during the training and validation phases. Considering the differences in magnitude and gradient scale among the different residual terms, the results for each branch loss in the figures are multiplied by their corresponding weight coefficients to accurately reflect the contribution of the physics constraints during the training process.
	
	Fig.~\ref{fig:loss_residuals}(a) shows the changing trends of the main physics residual loss terms during the training phase of the STWM-PINN model. The results indicate that all loss terms exhibit typical staged convergence characteristics. In the initial training stage (0–25,000 epochs), the seepage-mechanics residual loss $L_p$ and the elasto-plastic yield residual loss $L_y$ dominated the overall downward trend. This shows that the Darcy flow governing equation and the elasto-plastic constitutive constraints have a strong driving effect on the network weight adjustments in the early training, achieving rapid convergence driven by the large initial physical errors. In the middle stage (25,000–70,000 epochs), the losses entered a convergence period. The hydration-thermal diffusion residual loss $L_{wT}$ reached its minimum and remained relatively stable, indicating that the network has effectively fitted the diffusion-type physical mechanisms. In the late stage (70,000–100,000 epochs), the overall fluctuations of the residuals decreased. The variations of $L_p$ and $L_y$ were controlled within $\pm$0.8\% and $\pm$1.2\%, respectively, suggesting that the coupled physical fields have reached a state of high-consistency convergence. Notably, the hydration-thermal residual loss showed a slight upward trend in the final stage, implying that this branch is highly sensitive to the parameters.
	
	Fig.~\ref{fig:loss_residuals}(b) further verifies the generalization capability of the above training trends. The evolution of the total loss and the individual physics residual terms on the validation set was highly consistent with the training set. This highlights the robustness of the model to unseen data. However, the minimum values of the individual residual losses were generally elevated by 25\%–33\%. The minimum deviations for $L_p$ and $L_y$ were 0.2\% and 6.5\%, respectively. This validates the high stability of the Darcy-based seepage coupling term in spatial extrapolation. It also reveals that the elasto-plastic residual is more sensitive to changes in boundary conditions. In the late stage of training, the validation residual curves for $L_p$ and $L_y$ showed a slight oscillating upward trend, suggesting that the model may exhibit slight signs of overfitting under the strong physical constraints.
	
	As shown in Fig.~\ref{fig:loss_bcs}(a), the boundary and initial condition losses converge rapidly during the early iterations, decreasing from an initial order of $10^{-1}$ to within $10^{-3}$. This demonstrates the ability of the network to efficiently perceive and integrate boundary and initial information. The data fitting loss $L_{\text{data}}$ converged quickly in the early stage, decreasing from approximately unit magnitude to around $10^{-3}$, demonstrating the network’s good expressive power for the labeled samples. In the middle and late training stages, all three of these losses entered a slow-decline region with significantly reduced fluctuations, indicating that the various supervised branches of the network became increasingly coordinated. By the late stage of training, the total loss $L_{\text{total}}$ stabilized at the order of $10^{-3}$, reflecting stable convergence and effective residual control.
	
	Fig.~\ref{fig:loss_bcs}(b) further shows the evolution trend of the losses during the validation phase. The overall trend is consistent with the training phase. Both the boundary and initial condition losses converged within the range of $10^{-4}$ to $10^{-3}$ during validation. The data fitting loss was slightly higher than the training set result, but remained within a low-magnitude fluctuation range. The total loss stably oscillated between $10^{-3}$ and $10^{-2}$, with no obvious signs of divergence.
	
	\subsection{Model Prediction and Validation}
	\label{subsec:prediction}
	
	\subsubsection{Prediction Performance Validation}
\begin{figure*}[t]
		\centering
		\includegraphics[width=\textwidth]{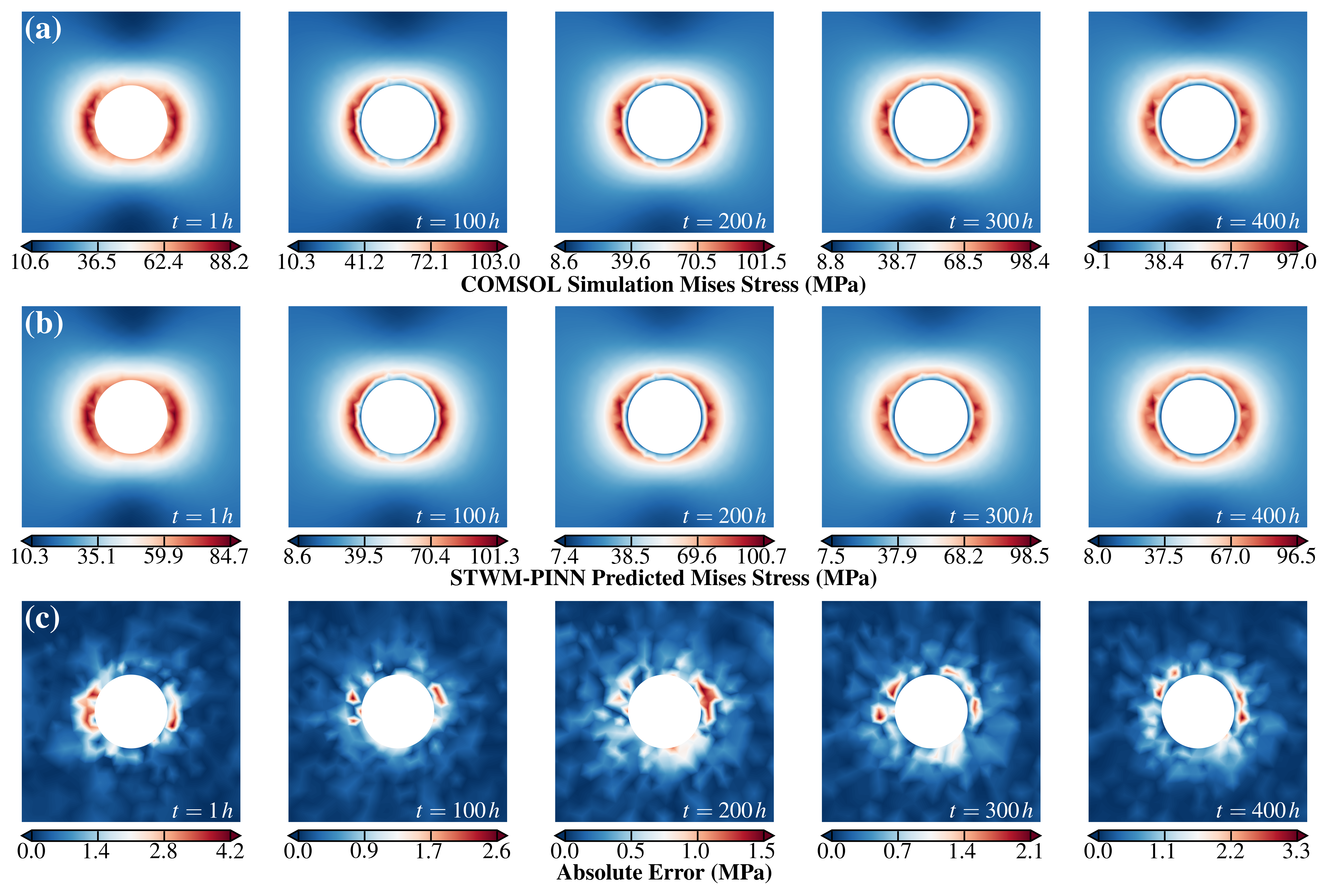}
		\vspace*{-10mm} 
		\caption{Spatial Distribution of Von Mises Stress Predictions and Error Field Evolution for the STWM-PINN Model.}
		\label{fig:pred_stress}
	\end{figure*}

\begin{figure*}[t]
		\centering
		\includegraphics[width=\textwidth]{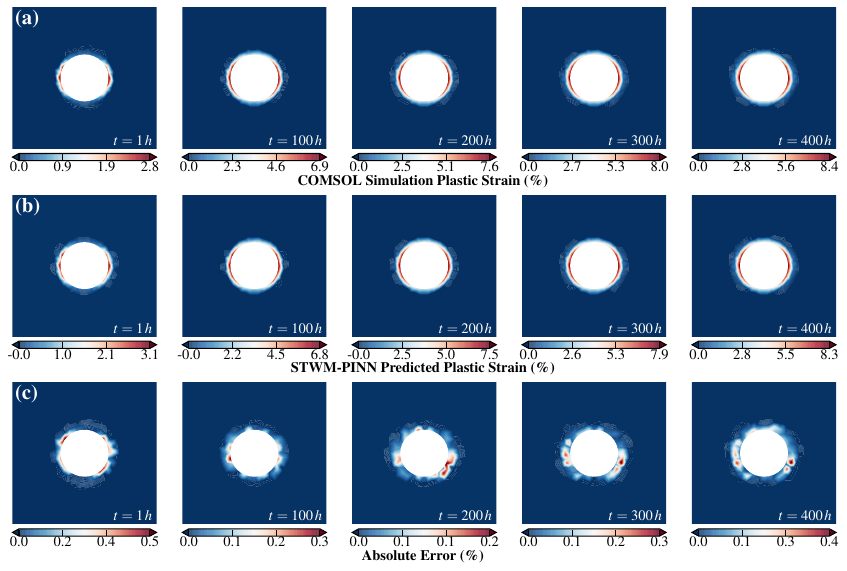}
		\vspace*{-10mm} 
		\caption{Spatial Distribution of Plastic Strain Predictions and Error Field Evolution for the STWM-PINN Model.}
		\label{fig:pred_strain}
	\end{figure*}

	Fig.~\ref{fig:pred_stress} and Fig.~\ref{fig:pred_strain} show the predictive performance of the STWM-PINN model for key geomechanical responses in the formation during the period $t = 1$~h to $t = 400$~h. Fig.~\ref{fig:pred_stress} focuses on the spatial distribution of Von Mises equivalent stress and its error evolution, whereas Fig.~\ref{fig:pred_strain} shows the prediction results for plastic strain and the change in its error over time.
	
	Fig.~\ref{fig:pred_stress}(a) shows the ground truth distribution of equivalent stress obtained from the COMSOL Multiphysics simulation. The Von Mises stress was primarily concentrated near the wellbore wall and diffused symmetrically in the radial direction. Over time, the main stress-controlled region gradually shifted toward the outer edge of the wellbore, while the overall distribution maintained a strong axisymmetric structure. Fig.~\ref{fig:pred_stress}(b) shows the prediction results from the STWM-PINN model. The model accurately predicted the main contours and evolution path of the high-stress regions, exhibiting good spatial matching, particularly in the stress concentration zones. It also effectively captured the temporal progression of stress concentration and diffusion. Fig.~\ref{fig:pred_stress}(c) shows the absolute error distribution between the ground truth and the predicted results. Regarding error evolution, the STWM-PINN controlled the prediction error to below 2~MPa for most of the evolution time. The error was mainly distributed in areas with large stress gradients and significant edge disturbances. In the early stage ($t = 1$h), the peak error exceeded 4~MPa and was concentrated in the stress transition zone near the side boundaries of the wellbore. Some areas, such as (X, Y) $\approx$ (-0.11, -0.01)~m, highly coincided with the maximum stress regions. Over time, the main error peak gradually migrated away from the main stress path. By $t = 400$~h, the maximum error value had decreased to approximately 3.26~MPa and was primarily located near the diagonally symmetric axis on the outer radial edge, showing a clear spatial shift.
	
	Fig.~\ref{fig:pred_strain}(a) shows the ground truth plastic strain field obtained from the COMSOL Multiphysics simulation. The plastic strain was mainly concentrated on both sides of the axisymmetric axis near the wellbore and exhibited a clear cumulative growth over time, showing strong spatio-temporal coupling. Fig.~\ref{fig:pred_strain}(b) shows the prediction results from the STWM-PINN model. The model accurately captured the main distribution and growth trends of the plastic strain, showing strong spatial prediction consistency, particularly within the axisymmetric domain.
	
	Fig.~\ref{fig:pred_strain}(c) then shows the absolute error distribution. It reveals the spatial distribution patterns of the prediction error of the model. Overall, the prediction error was concentrated in areas where the plastic strain gradient changes sharply or at boundary turning points. In the early stage ($t = 1$~h), the model showed a peak error of approximately 0.47\% in some strain concentration areas, such as (X, Y) $\approx$ (-0.09, 0.04)m, coinciding with the primary strain-controlled path. As the evolution progressed to the middle and late stages ($t \geq 200$~h), the region of maximum error gradually shifted toward the diagonally symmetric axis of the wellbore, showing an asymmetric spatial migration trend. The error magnitude reached 0.42\% at $t = 400$~h. Notably, from $t = 200$~h onward, the regions of maximum plastic strain and maximum error gradually separated. This indicates that the predictive stability of the STWM-PINN model in the main plastic strain-controlled areas improves as time progresses.
	
	\subsubsection{Extrapolation Generalization Validation}
\begin{figure*}[t]
		\centering
		\includegraphics[width=\textwidth]{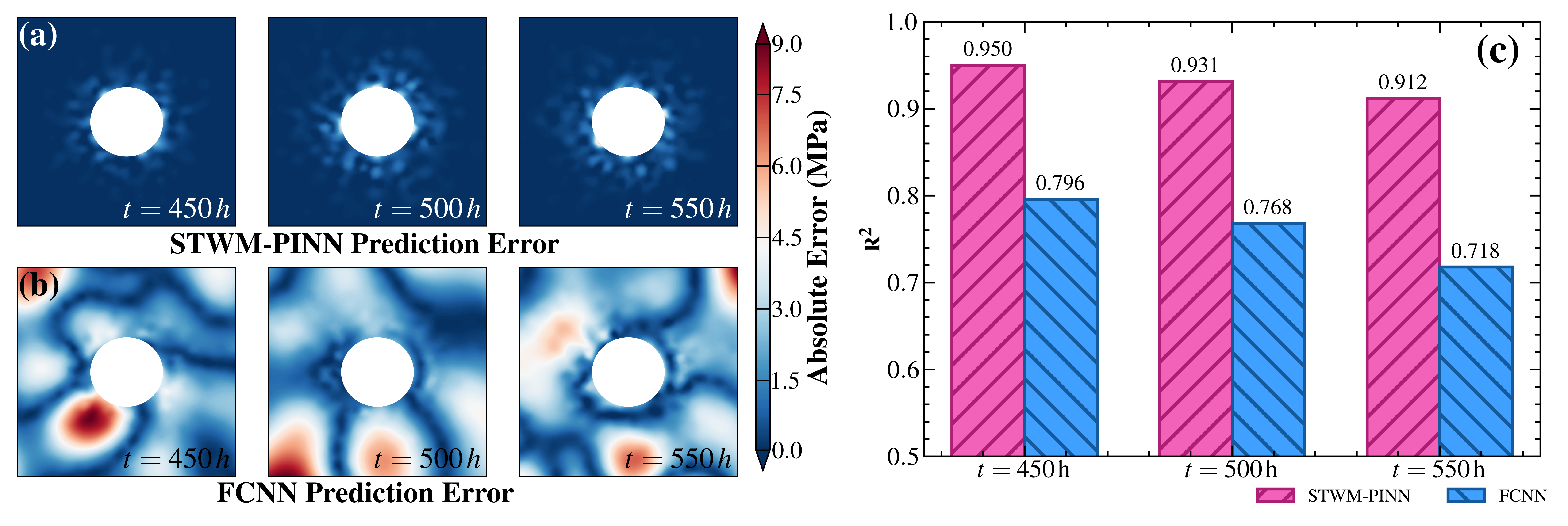}
		\vspace*{-10mm} 
		\caption{Comparison of Error Field Evolution and R² for the STWM-PINN and FCNN Models in Time Extrapolation.}
		\label{fig:extrapolation}
	\end{figure*}

	To further validate the generalization capability of the STWM-PINN model in the time dimension, three time points outside the training evolution window were selected: $t = 450$~h, $t = 500$~h, and $t = 550$~h. These time points were input into the fully trained STWM-PINN model to output predictions for the Von Mises equivalent stress. These results were simultaneously compared with the predictions generated by an FCNN under the same input conditions. This was done to evaluate the model’s physical consistency and spatial structure stability in a time-extrapolation scenario. Fig.~\ref{fig:extrapolation} shows the prediction results of the two models at the three aforementioned time points, systematically comparing the temporal evolution of the absolute error fields and the predictive accuracy.
	
	Fig.~\ref{fig:extrapolation}(a) shows the evolution of the absolute error field for the STWM-PINN model at $t = 450$~h, 500~h, and 550~h. In the time-extrapolation scenario, the prediction error of the STWM-PINN model was primarily concentrated in the areas of sharp stress gradients near the wellbore. The error remained within an acceptable range in the vast majority of the spatial domain and showed a stable and controllable growth trend over time. Fig.~\ref{fig:extrapolation}(b) shows the evolution of the absolute error field for the FCNN model at the same time points. The error of the FCNN model was more uniform across all regions and did not effectively concentrate in the areas with larger errors. Fig.~\ref{fig:extrapolation}(c) verifies the above description through a comparison of the $R^2$ values. The comparison results showed that the STWM-PINN model consistently exhibited a higher goodness-of-fit and stronger convergence consistency in predicting Von Mises equivalent stress, significantly outperforming the FCNN. This further proves the extrapolation capability and physical credibility of the STWM-PINN in long-term evolution prediction.
	
	The comprehensive analysis indicates that the STWM-PINN model provides a high-precision, high-stability, and high-physical-consistency solution for simulating the evolution of the open-hole wellbore under the coupled multiphysics disturbances of “seepage-thermal-hydration-mechanics” during long-term exposure. This lays a solid predictive foundation for carrying out intelligent control of drilling fluid parameters.
	
	\section{Development of a Deep Reinforcement Learning-Based Parameter Optimization Policy}
	\label{sec:drl}
	
	\subsection{Reinforcement Learning Environment Construction}
	\label{subsec:rl_env}
	
	\subsubsection{State Space Design}
	To enable intelligent control over the multiphysics evolution of an open-hole wellbore under long-term exposure, a state space with high physical expressiveness and spatial aggregation features must be constructed within the reinforcement learning framework. The core of the state space design is to accurately capture the spatial distribution patterns of key formation response variables during the drilling process. This design must also maintain strict consistency with the outputs of the STWM-PINN model to ensure the physical reliability of the environment’s input and the engineering interpretability of the state representation.
	
	The state variables consist of five core physical quantities output by the STWM-PINN model: pore pressure $p$, formation temperature $T$, water content $w$, Von Mises equivalent stress $\sigma_{\text{eq}}$, and plastic strain $\varepsilon_p$. To match the scale of the state space with the action space of the reinforcement learning policy, a spatial statistical aggregation method was adopted. The state space for each decision step $t$ is formed by the statistical features of a two-dimensional radial region:
	\begin{equation}
		s_t=\{\overline{p}(t),\overline{T}(t),\overline{w}(t),\overline{\sigma}_{\text{eq}}(t),\overline{\varepsilon}_p(t)\}
		\label{eq:23}
	\end{equation}
	where $s_t$ represents the state observation received by the policy network at the $t$-th decision step. The $t$ in parentheses denotes the decision step corresponding to the current combination of drilling fluid parameters, not the physical evolution time described by the STWM-PINN model. The overline on each state variable indicates a weighted average value within a specific spatial region. This approach significantly reduces the dimensionality of the state space and increases sensitivity to high-risk areas of the wellbore.
	
	The specific spatial aggregation region is defined as the near-wellbore boundary zone where the radial distance $r$ is in the range of [0.10, 0.13]~m and the vertical position satisfies $|y| \le 0.02$~m, that is:
	\begin{equation}
		\begin{cases}
			0.10 \le \sqrt{x^2+y^2} \le 0.13 \\
			|y| \le 0.02
		\end{cases}
		\label{eq:24}
	\end{equation}
	Within this boundary region, the state variables output by the STWM-PINN are spatially averaged. This process yields statistics that reflect the overall stability evolution characteristics of the boundary region. These statistics are used to represent the comprehensive impact of the current policy’s control parameters on the wellbore response. This state space design can accurately reflect wellbore boundary instability trends and is consistent with the action scale of the reinforcement learning policy network.
	
	On the time scale, the STWM-PINN model provided high-resolution spatio-temporal information for the physical evolution time $t' \in [0,200]$~h, with a step size of 1~h. However, the reinforcement learning policy did not explicitly model the time-series dependencies of the state variables. Each state $s_t$ reflected the spatial statistical performance of the formation response over the entire physical evolution period under the current set of drilling fluid parameters. This state served as the basis for the policy network’s decision. The subsequent physical state evolution trajectory was then used to evaluate the long-term control effect of the current action.
	
	\subsubsection{Action Space and Control Variable Definition}
	In a strongly coupled “seepage-thermal-hydration-mechanics” multiphysics system, the stability of the open-hole wellbore is highly dependent on the ability of the drilling fluid parameters to regulate the evolution of various field variables. To achieve direct control over the wellbore evolution path, the reinforcement learning agent must be able to continuously adjust key thermophysical properties of the drilling fluid. This allows for dynamic influence on the stress-strain response and instability risk of the wellbore without altering the constitutive properties of the formation.
	
	Drilling fluid density $\rho_w$, viscosity $\mu_0$, and temperature $T_m$ were selected as the three continuous control variables for the reinforcement learning environment. These correspond to the three core control mechanisms: pore pressure diffusion control, fluid resistance regulation, and thermal disturbance control. These variables have independent regulatory significance in the wellbore evolution process and are also embedded as explicit control variables in the construction of the PDE residuals within the STWM-PINN model. They affect the Biot-Darcy coupled equation, the energy conservation equation, and the hydration strain source term, respectively.
	
	Specifically, drilling fluid density $\rho_w$ regulates the fluid column pressure boundary condition and the seepage pressure differential field. This alters the rate and extent of pore pressure diffusion, playing a key role in the solid-fluid coupled stress propagation path. Viscosity $\mu_0$ determines the flow resistance coefficient in the Darcy flow process and also affects the contribution of the thermal convection term, acting as a bridge between the seepage and heat transfer paths. Drilling fluid temperature $T_m$ controls the interface temperature difference and the thermal boundary flux. This influences the intensity of thermal disturbance in the formation and has a significant impact on the thermo-hydration coupled field.
	
	To improve the convergence efficiency and physical interpretability of the policy network, the action vector was defined as a normalized relative adjustment:
	\begin{equation}
		\mathbf{a}_t=[\Delta\rho_w, \Delta\mu_0, \Delta T_m]^\top \in [-1,1]^3
		\label{eq:25}
	\end{equation}
	where $\mathbf{a}_t$ is the normalized action vector generated by the agent at the $t$-th policy step. $\Delta\rho_w$, $\Delta\mu_0$, and $\Delta T_m$ represent the standardized adjustments of the drilling fluid density, viscosity, and temperature relative to the median of their physically allowable ranges. To map the action vector to actual physical control variables, let $x_i \in \{\rho_w, \mu_0, T_m\}$ be any physical variable. The mapping relationship is as follows:
	\begin{equation}
		x_i = x_i^{\text{mid}} + \Delta x_i \cdot (x_i^{\max} - x_i^{\min})/2
		\label{eq:26}
	\end{equation}
	where $x_i^{\text{mid}}$, $x_i^{\min}$, and $x_i^{\max}$ represent the median, lower, and upper bounds of the allowable range for the variable $x_i$, respectively. $\Delta x_i$ represents the corresponding standardized adjustment value from the action. This formulation ensures the continuity of the action space and the differentiability required for policy gradient methods, while giving the policy output a clear engineering meaning. The value ranges for the three variables are defined as:
	\begin{equation}
		\begin{cases}
			\rho_w \in [980, 1080] \, \text{kg/m}^3 \\
			\mu_0 \in [0.025, 0.045] \, \text{Pa}\cdot\text{s} \\
			T_m \in [310, 350] \, \text{K}
		\end{cases}
		\label{eq:27}
	\end{equation}
	During training, the action $\mathbf{a}_t$ generated by the agent at each decision step $t$ was injected in real time as a control input into the STWM-PINN model and participated in the prediction of the state variable evolution in subsequent time steps.
	
	\subsubsection{Reward Function Design}
	To achieve effectiveness and physical consistency for the reinforcement learning policy in wellbore stability optimization, a composite reward function system was constructed. This reward system simultaneously accounts for engineering risk control, control continuity, and boundary feasibility. As the core feedback mechanism for the agent–environment interaction, the reward function’s expression is directly linked to the state variables predicted by the STWM-PINN model, while maintaining engineering interpretability. The designed immediate reward function $r_t$ consists of three parts:
	\begin{equation}
		\begin{split}
			r_t ={} & -\alpha_1 \tilde{P}_{f}^{\text{avg}}(t) - \alpha_2 \lVert \mathbf{a}_t-\mathbf{a}_{t-1} \rVert_2^2 \\
			& - \alpha_3 \sum_{i=1}^3 \mathbb{1}(x_t^{(i)} \notin [x_{\min}^{(i)}, x_{\max}^{(i)}])
		\end{split}
		\label{eq:28}
	\end{equation}
	where $r_t$ represents the immediate reward at step $t$; $\tilde{P}_{f}^{\text{avg}}(t)$ represents the average wellbore instability risk over the evolution time $t' \in [0,200]$~h; $\mathbf{a}_t$ represents the current normalized action vector; $x_t^{(i)}$ represents the $i$-th physical variable mapped from the current action $\mathbf{a}_t$; $\mathbb{1}(\cdot)$ represents the indicator function; and the weights $\alpha_1$, $\alpha_2$, $\alpha_3$ are used to control the relative contributions of the instability risk penalty, the action smoothness constraint, and the physical boundary violation penalty to the total reward.
	
	The first term, $\tilde{P}_{f}^{\text{avg}}(t)$, is the core penalty term in the policy optimization process. It represents the average instability risk of the wellbore over the entire evolution cycle under the influence of the current action $\mathbf{a}_t$. To enhance the physical consistency and differentiability of the risk assessment term, the degradation relationships defined by the hydration-thermal-mechanical coupling mechanism from Eq.~(\ref{eq:7}) were substituted. The cohesion $c(w)$ and friction angle $\phi(w)$, driven by hydration evolution, were mapped to the parameters $a$ and $k$ of the Drucker-Prager yield criterion. This mapping relationship is detailed in Eq.~(\ref{eq:17}). Based on this, and combined with the formation’s shear response and constitutive relationship, the expressions for the equivalent yield stress and strain thresholds can be obtained:
	\begin{equation}
		\begin{cases}
			\sigma_y = k/a \\
			\varepsilon_y = \sigma_y/E(w) + 0.002
		\end{cases}
		\label{eq:29}
	\end{equation}
	where $\sigma_y$ represents the equivalent yield stress under hydration-induced weakening; $\varepsilon_y$ represents the yield strain threshold; 0.002 is an empirical initial strain reflecting the onset of irreversible deformation; and $E(w)$ represents the degraded elastic modulus under the influence of hydration.
	
	Based on the predictions from the STWM-PINN model in the wellbore spatial domain, the equivalent stress $\sigma_{\text{eq}}$ and plastic strain $\varepsilon_p$ were extracted. They were then normalized as $\sigma_{\text{eq}}/\sigma_y$ and $\varepsilon_p/\varepsilon_y$, respectively. To improve the smoothness and differentiability of the risk assessment function, a logistic mapping function is introduced:
	\begin{equation}
		L(x)=\frac{1}{1+\exp(-b(x-x_0))}
		\label{eq:30}
	\end{equation}
	where $b=12$ and $x_0=1$. This ensures a sensitive response and a clear gradient within the critical yield range. The instantaneous failure probability function constructed accordingly is:
	\begin{equation}
		P_f(t') = 0.4 \cdot L(\sigma_{\text{eq}}/\sigma_y) + 0.6 \cdot L(\varepsilon_p/\varepsilon_y)
		\label{eq:31}
	\end{equation}
	Then, by time-averaging the instantaneous failure probability $P_f(t')$ over the interval $t' \in [0,200]$~h, the average wellbore failure risk corresponding to the current policy action is defined as:
	\begin{equation}
		\tilde{P}_{f}^{\text{avg}}(t) = \frac{1}{201} \sum_{t'=0}^{200} P_f(t'; \mathbf{a}_t)
		\label{eq:32}
	\end{equation}
	where the denominator 201 represents the number of time steps in the evolution interval $[0,200]$~h. This is used to normalize the failure probability sequence output by the STWM-PINN model, reflecting the overall impact of the current control policy on wellbore stability throughout the entire physical evolution cycle.
	
	The second term is the action continuity penalty. It quantifies the Euclidean distance between the current action $\mathbf{a}_t$ and the previous action $\mathbf{a}_{t-1}$. This term is intended to limit drastic fluctuations in the control of drilling fluid parameters. It prevents numerical instability or nonlinear amplification effects in field operations that could be caused by significant adjustments. This enhances the temporal consistency and engineering feasibility of the policy, consistent with the principle of gradual adjustment in practical well control operations.
	
	The third term is the physical boundary constraint. It uses an indicator function to determine whether the current mapped variable $x_t^{(i)}$ has exceeded its preset physically allowable range $[x_{\min}^{(i)}, x_{\max}^{(i)}]$. If a violation occurs, an explicit negative penalty is applied. This ensures that the generated actions are feasible at the physical, numerical, and engineering levels.
	
	\subsection{Policy Optimization Method}
	\label{subsec:policy_opt}
\begin{figure*}[t]
		\centering
		\includegraphics[width=0.9\textwidth]{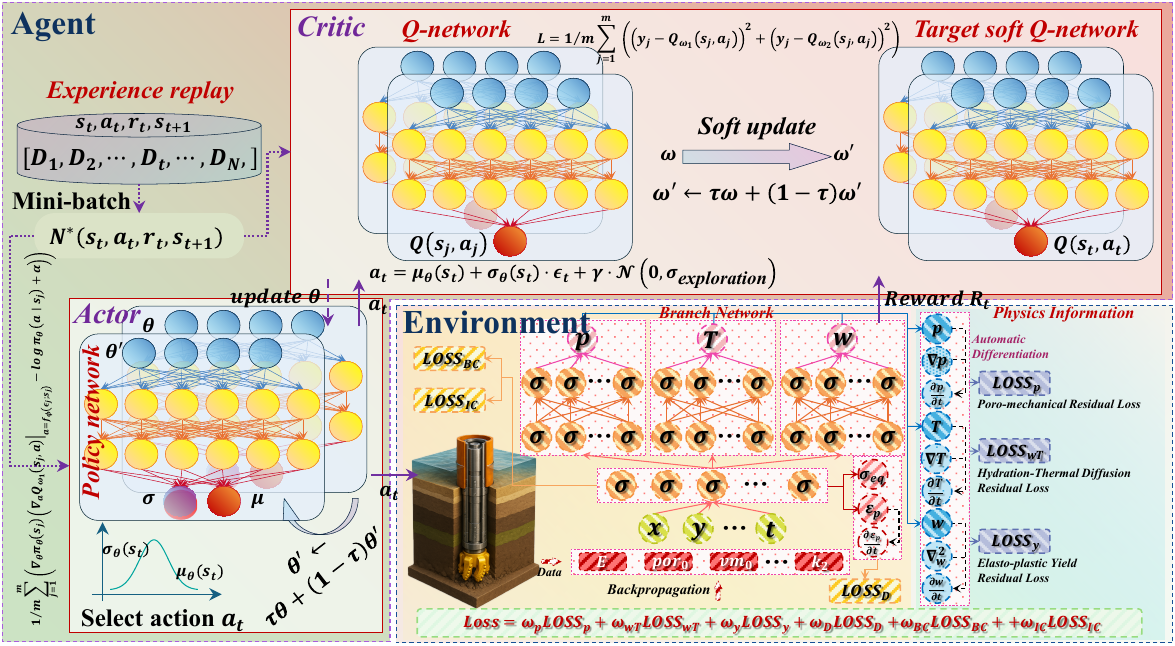}
		\vspace*{-3mm} 
		\caption{Architecture of the DN-SAC Policy Optimization Algorithm.}
		\label{fig:dnsac_arch}
	\end{figure*}

	To achieve optimal control of the drilling fluid parameters (density, viscosity, and temperature) within a continuous action space, a previously published algorithm, the Double-Noise Soft Actor-Critic (DN-SAC), was introduced as the primary policy optimization structure \citep{ref40}. It was integrated into the STWM-PINN multiphysics coupled environment. This algorithm is based on maximum entropy reinforcement learning. It can explore efficiently and converge stably in complex, non-stationary state spaces. Its robustness and convergence performance have been demonstrated in previous drilling parameter optimization problems. The algorithm architecture is shown in Fig.~\ref{fig:dnsac_arch}.
	
	Regarding implementation, the DN-SAC algorithm inherits the twin Q-network structure and soft parameter update mechanism, while employing a dual-noise design of policy noise and exploration noise \citep{ref41}. The former is introduced through a reparameterization mechanism to ensure the differentiability of policy updates and the stability of gradient propagation. The latter expands the action sampling space through an independently controllable, fixed Gaussian perturbation, enhancing the policy’s global optimization capability and convergence robustness for non-convex objectives. The experience replay module uses a prioritized experience replay mechanism \citep{ref42}. It dynamically adjusts sampling probabilities based on the temporal difference error to optimize the training efficiency for critical states. During the interaction process, the policy network takes the state variables generated by the STWM-PINN as input and outputs the optimized control actions for the drilling fluid parameters. It achieves closed-loop dynamic regulation in the physical response field through state updates and reward feedback.
	
	The relevant algorithm framework, network structure, and update procedures have been described in detail in previous work. Based on that, the core structure and hyperparameter configuration are retained. They are adapted and adjusted according to the multiphysics response characteristics of wellbore stability. The training configuration is shown in Table~\ref{tab:dnsac_hyperparams}.
	
	\begin{table*}[t]
		\centering
		\caption{Training Hyperparameter Configuration for the DN-SAC Algorithm.}
		\label{tab:dnsac_hyperparams}
		\begin{tabular}{llc}
			\hline
			\textbf{Symbol} & \textbf{Description} & \textbf{Value} \\ \hline
			$L$ & Number of hidden layers for Policy \& Q Networks & 3 \\
			$N_h$ & Neurons per layer & [256, 128, 64] \\
			$\sigma(\cdot)$ & Activation function & GELU \\
			$\eta_0$ & Initial learning rate & $3.0\times10^{-4}$ \\
			$\alpha$ & Entropy regularization temperature coefficient & 0.2 \\
			$\gamma$ & Reward discount factor & 0.99 \\
			$\tau$ & Target network soft update coefficient & $5.0\times10^{-3}$ \\
			$B$ & Batch size & 256 \\
			$N_{\text{explore}}$ & Initial pure exploration steps & 5000 \\
			$N_{\text{replay}}$ & Prioritized experience replay buffer capacity & $1.0\times10^5$ \\
			$N_{\text{train}}$ & Total training rounds & $1.0\times10^5$ \\
			$N_{\text{patience}}$ & Early Stopping patience rounds & 3000 \\
			$\delta_{\text{stop}}$ & Early Stopping threshold & $1.0\times10^{-6}$ \\ \hline
		\end{tabular}
	\end{table*}
	
	During training, the optimization of the DN-SAC policy is driven by dual objective functions. The first is to minimize the average Temporal Difference (TD) error between the twin Q-networks ($Q_{\omega_1}$ and $Q_{\omega_2}$), thereby improving the accuracy of the state-action return estimation. The second is to maximize the policy entropy regularization term, which encourages the generation of a diverse and robust action distribution. To unify the metrics and improve interpretability, the TD losses of the two Q-networks are averaged during training to serve as the Critic Loss. This reflects the overall fitting quality of the value function estimation process. The Actor Loss, by contrast, is composed of the minimum policy entropy objective and a Q-value guidance term. This measures the ability of the policy network to fit the optimal action distribution while maintaining the benefits of entropy-driven exploration.
	
	\begin{figure*}[t]
		\centering
		\includegraphics[width=\textwidth]{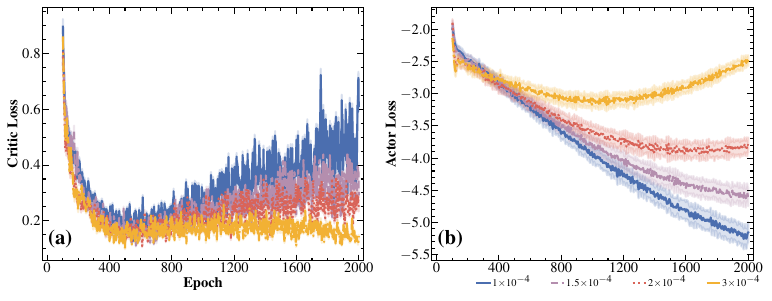}
		\vspace*{-10mm} 
		\caption{Comparison of Actor-Critic Loss Evolution Trends for the DN-SAC Algorithm under Different Learning Rate Configurations.}
		\label{fig:ac_loss}
	\end{figure*}
	
	Fig.~\ref{fig:ac_loss} shows the evolution of the Critic Loss and Actor Loss with training steps under different learning rate configurations, highlighting the substantial impact of the learning rate on the policy’s convergence path and optimization stability. Fig.~\ref{fig:ac_loss}(a) shows the change in the Critic Loss curve, which reflects the convergence of the mean squared error during the return function estimation process of the twin Q-networks. Overall, all configurations generally exhibited staged training characteristics. They showed a significant downward trend, particularly in the early stage (0–500 timesteps), indicating that the Q-networks rapidly adapted to the return estimation task under the initial policy. In the middle and late stages (500–1200 timesteps), differentiation between the configurations gradually appeared. The $1.5\times10^{-4}$ learning rate performed best in terms of convergence speed and curve stability. Its loss value rapidly approached a low level and remained in a stable range, reflecting its good generalization ability in value function estimation. Conversely, the $1\times10^{-4}$ learning rate showed a stable initial decline, but its fluctuations increased in the later stage, possibly due to insufficient gradient update magnitude. The $3\times10^{-4}$ learning rate converged rapidly at first but showed increased loss fluctuations and a slight rebound in the later stage. This suggests an imbalance in the update rhythm between the policy and Q-networks, which could lead to overfitting or oscillation in the return estimation.
	
	Fig.~\ref{fig:ac_loss}(b) further analyzes the evolution of the policy network’s Actor Loss. Since this loss term is composed of a negative entropy regularization term and a Q-value minimization guidance term, a lower value indicates better performance of the current policy in generating high-value, high-entropy action distributions. Therefore, the Actor Loss is typically negative and continuously decreases during the training process. With the $1.5\times10^{-4}$ configuration, the Actor Loss shows a significant decrease, reaching a minimum of -4.65, and maintains a stable convergence state in the later stage. This indicates that this learning rate can effectively balance policy exploration and goal-orientation, avoiding issues of policy degradation or oscillation. Conversely, the $1\times10^{-4}$ curve achieves a deeper minimum (-5.28) but shows greater overall fluctuations, reflecting risks of insufficient policy updates. The $3\times10^{-4}$ curve exhibits a clear rebound between the 1200–2000 timesteps, suggesting that a high learning rate may lead to excessive policy updates, a disordered entropy structure, and even performance degradation.
	
	\subsection{Validation of Optimization Performance}
	\label{subsec:opt_validation}
	
	\subsubsection{Analysis of Control Parameter Characteristics}
	The formation in the 832~m well section transitions from an overlying loose muddy layer to a dense mudstone. The porosity decreased from 0.065 to 0.049, and the permeability was significantly reduced, limiting the invasion of drilling fluid filtrate. The cohesion increased from 0.74~MPa to 1.63~MPa, reflecting an enhancement in shear strength and structural integrity. The internal friction angle remained stable between 24.63° and 24.67°. Overall, this section consists of a dense mudstone with good mechanical properties and a stable in-situ stress environment, possessing high natural wellbore stability. Despite this, the time-dependent degradation effects driven by multiphysics coupling during long-term exposure are still the key factors leading to its eventual instability, providing a clear objective and challenge for parameter optimization.
	
	Against this dense mudstone background, a comparison of the performance of different reinforcement learning strategies in wellbore stability control was conducted. Four representative algorithms were selected: Deep Deterministic Policy Gradient (DDPG), Twin Delayed DDPG (TD3), Soft Actor-Critic (SAC), and the Double-Noise SAC (DN-SAC) algorithm used in this study \citep{ref41, ref43, ref44}. All policies were deployed under the same geological environment and initial conditions. The objective was to jointly optimize the drilling fluid density, viscosity, and temperature to reduce plastic strain and enhance stability in the open-hole boundary region.
	
	\begin{figure*}[t]
		\centering
		\includegraphics[width=\textwidth]{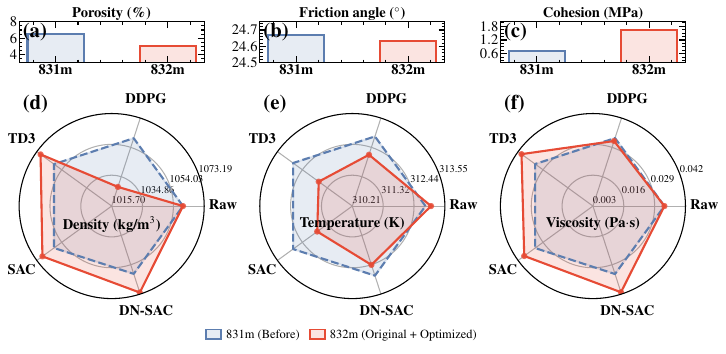}
		\vspace*{-10mm} 
		\caption{Analysis and Comparison of Rock Properties and Multi-Algorithm Optimized Drilling Fluid Parameters for the 832~m Well Section.}
		\label{fig:param_analysis}
	\end{figure*}
	
	As shown in Fig.~\ref{fig:param_analysis}, the original parameters for this comparative analysis were set as a density of 1060~kg/m³, a viscosity of 0.0334~Pa·s, and a temperature of 313.05~K. The DDPG policy, being risk-averse, recommended a density of 1028.2~kg/m³ and a viscosity of 0.0321~Pa·s. This reflects a conservative control characteristic suitable for conditions with low formation disturbance. The TD3 and SAC policies increased the density to 1070.3~kg/m³ and 1068.9~kg/m³, and the viscosity to 0.0407~Pa·s and 0.0392~Pa·s, respectively. This further enhances the inhibition of filtrate invasion and the support effect on the shale, reflecting a control strategy centered on high-pressure fluid column support. In comparison, the DN-SAC policy further strengthened the fluid column support mechanism. It recommended a density of 1072.2~kg/m³ and a viscosity of 0.0417~Pa·s, while also slightly reducing the temperature to 312.44~K to synergistically weaken thermal-induced stress fluctuations.
	
	\subsubsection{Comparison of Failure Probability Evolution Before and After Optimization}
	To evaluate the effectiveness of the DN-SAC reinforcement learning policy in enhancing the stability of the open-hole wellbore, a predicted failure probability distribution field for the 832m section was constructed based on the STWM-PINN model. Key nodes were selected for comparative analysis at 25-hour intervals between 100 and 200~h. A comparative analysis of the wellbore state evolution was conducted for the original parameter case and the cases controlled by the four optimization algorithms (DDPG, TD3, SAC, DN-SAC).
	
	\begin{table*}[t]
		\centering
		\caption{Comparison of Time-Series Evolution and Reduction Magnitude of Wellbore Failure Probability under Different Reinforcement Learning Algorithm Optimizations.}
		\label{tab:failure_prob}
		\begin{tabular}{lccccc}
			\hline
			\textbf{Evolution Time (h)} & \textbf{Original} & \textbf{DDPG} & \textbf{TD3} & \textbf{SAC} & \textbf{DN-SAC} \\ \hline
			100 & 0.4187 & 0.4102 & 0.3815 & 0.3693 & 0.3550 \\
			125 & 0.4535 & 0.4457 & 0.4192 & 0.4154 & 0.3940 \\
			150 & 0.4831 & 0.4761 & 0.4518 & 0.4441 & 0.4271 \\
			175 & 0.4880 & 0.4811 & 0.4573 & 0.4479 & 0.4330 \\
			200 & 0.5073 & 0.5012 & 0.4795 & 0.4689 & 0.4558 \\
			Avg. Reduction & — & 1.56\% & 6.94\% & 8.81\% & 12.27\% \\ \hline
		\end{tabular}
	\end{table*}
	
	\begin{figure*}[t]
		\centering
		\includegraphics[width=\textwidth]{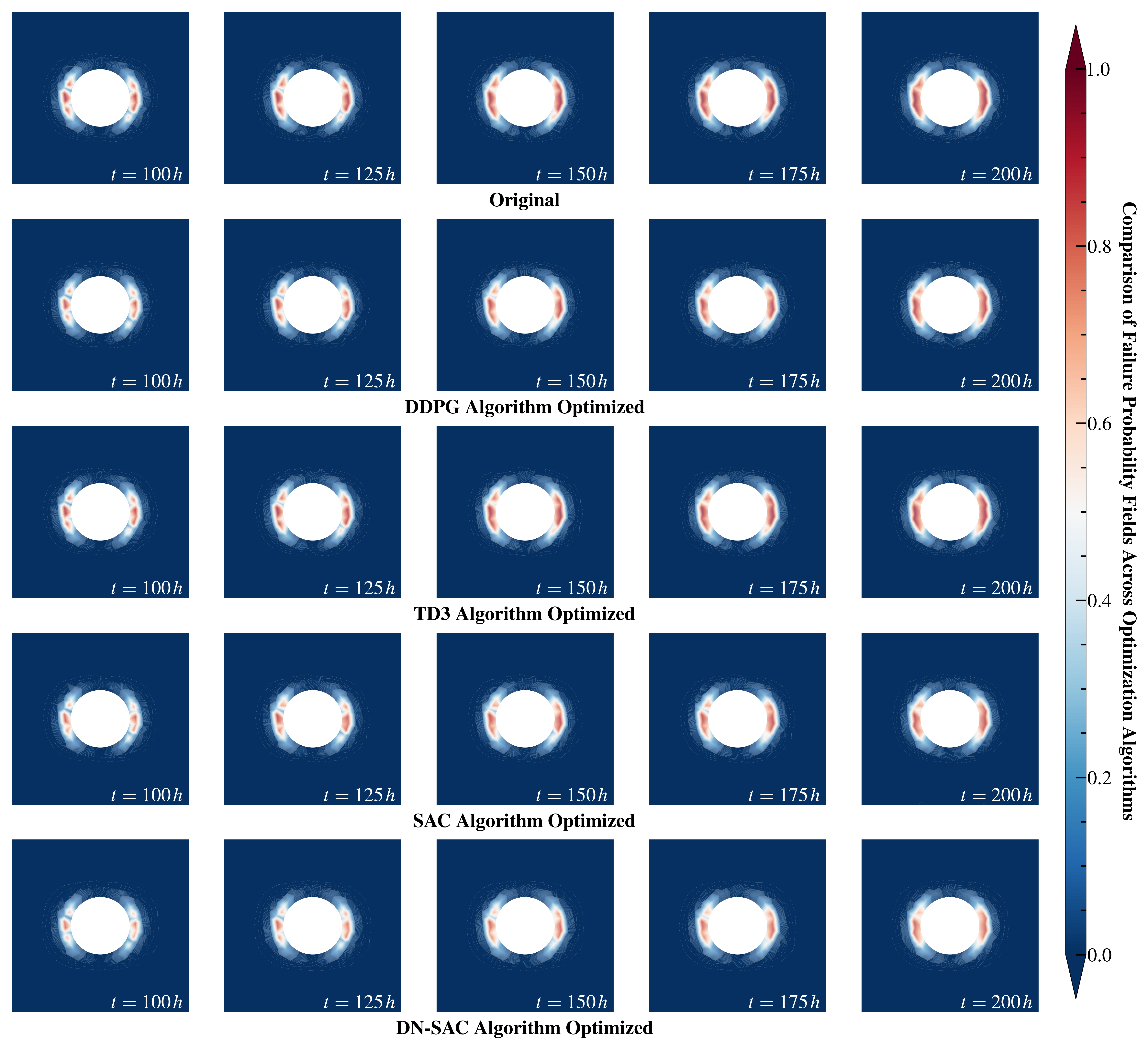}
		\vspace*{-10mm} 
		\caption{Comparison of the Time-Series Evolution of Wellbore Failure Probability Fields under Different Optimization Strategies.}
		\label{fig:failure_prob_fields}
	\end{figure*}
	
	Fig.~\ref{fig:failure_prob_fields} and Table~\ref{tab:failure_prob} jointly display the time-series evolution of wellbore failure probability under the five strategies. Fig.~\ref{fig:failure_prob_fields} presents the spatial distribution images of the probability fields, while Table~\ref{tab:failure_prob} provides the quantitative statistical results at key time nodes. Under the original conditions, the failure probability significantly increased with evolution time, from 0.4187 at $t = 100$~h to 0.5073 at $t = 200$~h. The high-risk area was mainly concentrated at the near-wellbore boundary of the open-hole section with $r \ge 0.1$~m. This reflects a coupled instability process driven by hydration, thermal disturbance, and stress migration under long-term exposure. The optimization algorithms significantly improved this trend, with each policy achieving effective suppression of the failure probability at different time points. The final failure probability under DDPG control was 0.5012. TD3 and SAC reduced it to 0.4795 and 0.4689, respectively. DN-SAC achieved the optimal control, reducing the value to 0.4558 by $t = 200$~h. Calculating the average reduction, DDPG, TD3, SAC, and DN-SAC achieved failure probability reductions of 1.56\%, 6.94\%, 8.81\%, and 12.27\%, respectively. Notably, DN-SAC consistently maintained a significant ability to suppress the instability risk at the wellbore boundary throughout the entire evolution. It demonstrated superior control stability and convergence consistency compared to the other policies. This performance improvement is attributed to the dual-noise exploration mechanism adopted by the DN-SAC algorithm. This mechanism cleverly combines two complementary noise sources. The intrinsic stochasticity introduced through policy reparameterization, inherited from the SAC framework, is mainly responsible for ensuring the stability of policy gradient calculations and encouraging fine-grained exploration around the current optimal action. Conversely, the external independent Gaussian noise, unique to the DN-SAC algorithm, injects stronger, policy-independent perturbations into the action space, effectively preventing the policy from prematurely converging to a local optimum. The synergistic effect of these two noises—the former ensuring stable convergence and the latter promoting broad exploration—significantly enhances the algorithm’s global optimization capability in a complex parameter space, enabling it to discover more robust and effective control policies.
	
	\subsubsection{Spatio-temporal Analysis of Boundary Stability Improvement}
	To analyze the role of different control policies in enhancing the stability of the open-hole wellbore boundary region, a comparison between the DN-SAC optimization policy and the original parameter case was performed. The time-series evolution and spatial distribution improvement of the failure probability at different evolution times were evaluated.
	
	\begin{figure*}[t]
		\centering
		\includegraphics[width=\textwidth]{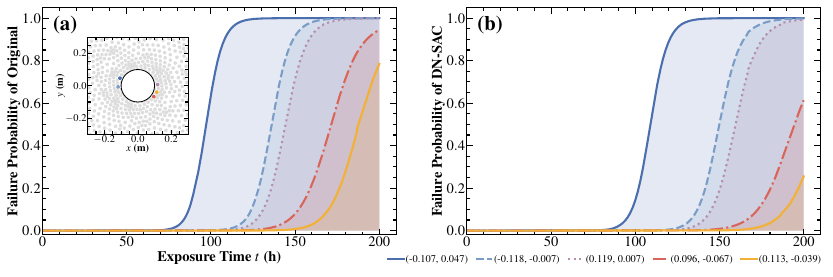}
		\vspace*{-10mm} 
		\caption{Time-Series Evolution of Failure Probability at Key Wellbore Points under Original and DN-SAC Optimized Policies.}
		\label{fig:key_points_prob}
	\end{figure*}
	
	In terms of temporal evolution, Fig.~\ref{fig:key_points_prob}(a) shows the failure probability evolution of the five highest-risk key points within the wellbore boundary region at times $t = 100, 125, 150, 175, 200$~h. The results show that under the original parameter conditions, the failure probability continuously increases with time, with the rate of increase accelerating after $t = 125$~h, showing a clear trend of nonlinear, accelerated degradation. For example, at the point (-0.107, 0.047), the failure probability rapidly increases. This reflects that in regions of highly water-sensitive mudstone, local instability is prone to occur under the coupled effects of thermal, hydraulic, and stress fields, becoming a high-risk exposure point. In contrast, Fig.~\ref{fig:key_points_prob}(b) shows that under the control of the DN-SAC optimization policy, the growth trend of the failure probability is significantly slowed. The failure probability at some key points is markedly reduced, showing an improvement in temporal stability after control. For instance, at $t = 150$~h, the failure probability at the point (-0.118, -0.007) decreased from 0.91570 before optimization to 0.50294, significantly delaying the onset of the high-risk state. This demonstrates the ability of the policy to mitigate the instability trend during the evolution process.
	
	\begin{figure*}[t]
		\centering
		\includegraphics[width=\textwidth]{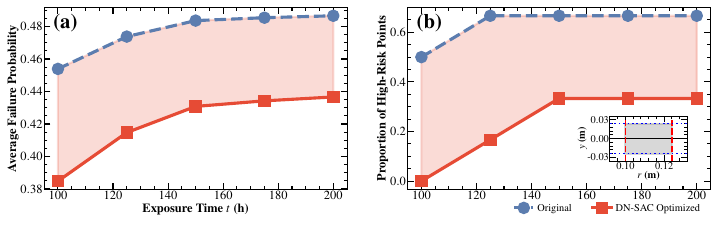}
		\vspace*{-10mm} 
		\caption{Improvement Effect of the DN-SAC Optimization Policy on the Statistical Features of Wellbore Instability in the Boundary Region.}
		\label{fig:stat_features_prob}
	\end{figure*}
	
	Regarding spatial distribution, Fig.~\ref{fig:stat_features_prob}(a) shows the trend of the average failure probability in the open-hole wellbore boundary region at different times. The results indicate that under the original conditions, the average failure probability steadily increases with time, and the cumulative instability risk is significantly enhanced. Fig.~\ref{fig:stat_features_prob}(b) shows the growth trend of the proportion of high-risk points ($P_f>0.5$) with evolution time, which also shows a time-dependent increase in high-risk exposure. Conversely, under DN-SAC optimal control, both of these metrics were effectively suppressed. The optimization policy achieved an average failure probability reduction of about 11.9\% and a high-risk point proportion reduction of about 65\% over the entire evolution period. It effectively weakened the instability trend that accumulates over time, enhancing the stability and safety margin of the boundary region under long-term exposure.
	
	\begin{figure*}[t]
		\centering
		\includegraphics[width=\textwidth]{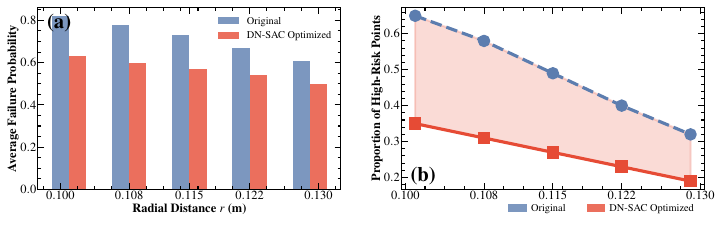}
		\vspace*{-10mm} 
		\caption{Analysis of the Improvement Effect of the DN-SAC Policy on the Radial Distribution of Failure Risk in the Boundary Region.}
		\label{fig:radial_prob}
	\end{figure*}
	
	Furthermore, to evaluate the improvement in the boundary region along the radial spatial scale, Fig.~\ref{fig:radial_prob}(a) compares the average failure probability distribution in different radial intervals of the wellbore boundary region (0.10 ~ 0.13~m) at $t = 200$~h. Fig.~\ref{fig:radial_prob}(b) shows the corresponding distribution of the proportion of high-risk points in each radial interval. The results show that the DN-SAC optimization policy achieved effective risk suppression in all radial intervals. For example, in the 0.100 ~ 0.106~m interval, the average failure probability was reduced by 23.2\%, and the proportion of high-risk points decreased by 46.2\%. In the 0.124 ~ 0.130~m interval, the average failure probability was reduced by 18.0\%, and the proportion of high-risk points decreased by 40.6\%. Overall, the DN-SAC optimization policy achieved a control effect of a 21.1\% reduction in average failure probability and a 44.1\% reduction in the proportion of high-risk points across the entire radial range. This verifies its stable control performance and coverage capability in the spatial domain.
	
	\subsection{Case Analysis}
	\label{subsec:case_analysis}
	To validate the effectiveness of the DN-SAC policy in extending the tolerable exposure time of the open-hole wellbore in complex formation conditions through the optimization of drilling fluid parameters, a case analysis was conducted on an 809–910~m open-hole section of a well in the Bohai Caofeidian block, China. This section is in the third drilling phase, with the casing from the second phase set at a depth of 808~m. The wellbore is exposed to the drilling fluid environment for a long time without casing support. Its stability is comprehensively affected by the coupled multiphysics disturbances of pore pressure diffusion, the geothermal gradient, hydration, and the mechanical heterogeneity of the formation. Based on the wellbore stability evolution prediction method constructed with the STWM-PINN model, the DN-SAC policy was used to solve for the optimal combination of drilling fluid parameters for a given set of formation properties. Specifically, for the formation properties at different depths within the interval, the DN-SAC policy outputs a single set of recommended drilling fluid density, viscosity, and temperature values, which aim to maximize the stability over the entire period. The time to first instability is defined as the evolution time point when the average failure probability first exceeds 0.5 within a near-wellbore window.
	
	\begin{figure*}[t]
		\centering
		\includegraphics[width=\textwidth]{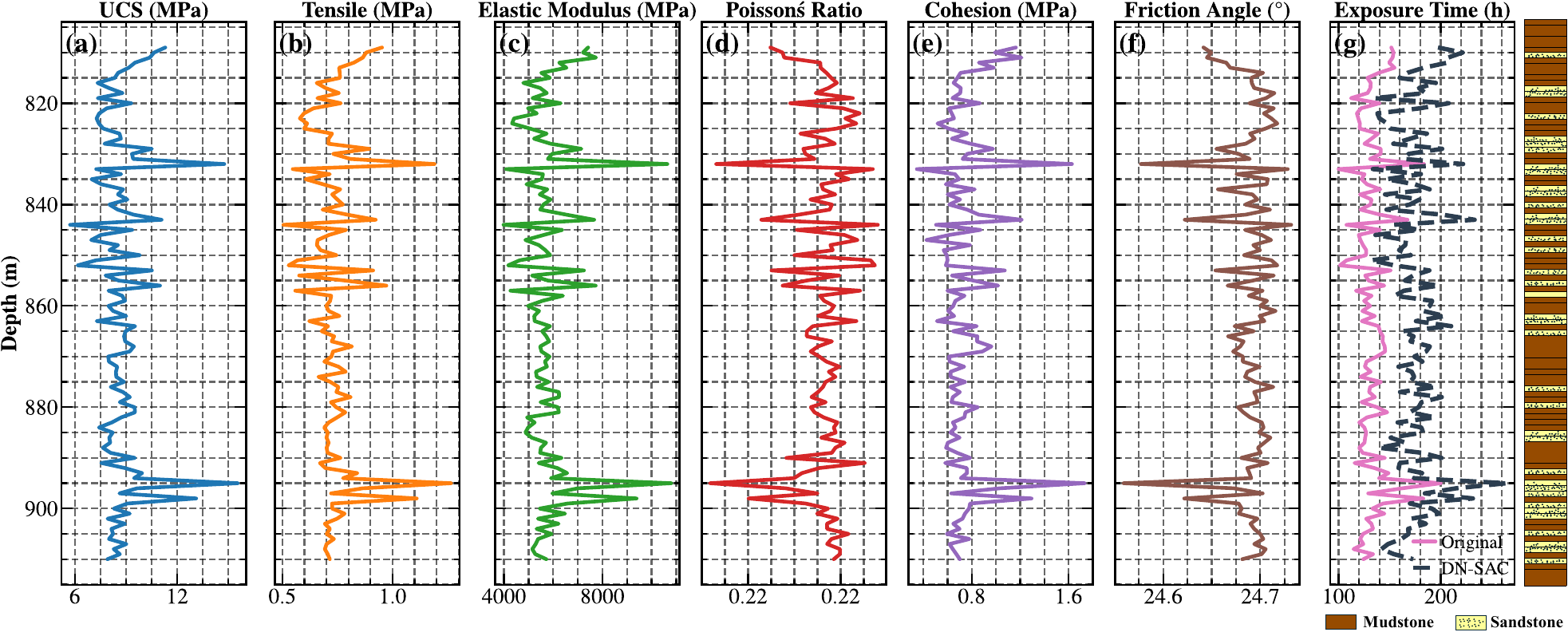}
		\vspace*{-10mm} 
		\caption{Case Validation of the DN-SAC Policy in Extending Open-Hole Exposure Time.}
		\label{fig:case_validation}
	\end{figure*}
	
	As shown in Fig.~\ref{fig:case_validation}, the results indicate that the DN-SAC policy achieved a significant improvement in open-hole wellbore stability in this section. An enhanced stability effect was observed across the entire depth range. Compared to the initial control state, the time to first instability was delayed by an average of 32.33\% after optimization. The maximum delay reached 53.35\% (at a depth of 859~m), and the minimum delay was 13.69\% (at a depth of 846~m). This demonstrates the policy’s adaptability and control robustness across different structural and lithological backgrounds. Particularly in key layers with high original failure probability and weak stability margins, the DN-SAC policy was able to effectively delay the onset of local instability, significantly enhancing the ability to maintain wellbore integrity.
	
	Further statistical analysis shows that in low-strength formations with cohesion below 1~MPa (93 layers), the DN-SAC policy can delay the time to first instability by an average of 32.46\%. In low-stiffness formations with an elastic modulus below 6000~MPa (75 layers), the delay is 32.68\%. In high-risk layers with an initial failure probability above 0.247 (19 layers), an average improvement of 32.62\% can also be obtained. These results verify the wide applicability and stability-enhancing control effect of the policy in various types of weakly stable formations. It exhibits excellent optimization search capabilities, particularly in sections sensitive to stress boundaries and significant mechanical disturbances, significantly extending the tolerable timescale of wellbore stability evolution.
	
	The comprehensive analysis shows that the DN-SAC optimization policy exhibits excellent capabilities in delaying the temporal evolution and improving the spatial distribution of failure risk in the open-hole wellbore boundary region. Through effective identification of high-risk points and targeted parameter control, the policy can achieve significant suppression of local failure risk, extend the safe exposure window of the boundary region, and enhance wellbore stability. It provides a generalizable intelligent optimization and control scheme for drilling operations in complex formations.
	
	\section{Conclusions}
	\label{sec:conclusions}
	
	\begin{enumerate}
		\item A high-fidelity multiphysics coupled numerical model was constructed by integrating Biot’s modified momentum conservation equation, Darcy’s flow equation, the non-steady-state heat transfer equation, and the hydration diffusion governing equation. The model accurately captured the radial diffusion patterns of the pore pressure and temperature fields under continuous drilling fluid invasion. It also captured the localized accumulation and spatially limited characteristics of the water content. The work demonstrated the model’s capability for fine-grained characterization of the synergistic effects of the “seepage-thermal-hydration-mechanics” four-field coupling and the resulting stress field redistribution and plastic zone evolution under long-term exposure conditions.
		
		\item The STWM-PINN architecture was proposed to achieve physically consistent predictions of the open-hole wellbore state through a multi-branch, physics-constrained neural network. By embedding governing equation residuals and boundary/initial constraints via automatic differentiation, the model effectively predicted the evolution of pore pressure, temperature, water content, Von Mises equivalent stress, and plastic strain. This was achieved under the supervision of a finite set of high-fidelity data points. The model also demonstrated good temporal extrapolation generalization ability, laying the surrogate model foundation for efficient wellbore state prediction and parameter optimization.
		
		\item An optimized decision-making methodology to minimize instability probability was proposed and validated. This method constructed a composite reward function that considered control smoothness and physical boundary constraints. Combined with the DN-SAC reinforcement learning algorithm, it optimized drilling fluid density, viscosity, and temperature. Case validation showed that this method can significantly reduce the probability of open-hole wellbore instability and extend the stable exposure time under different structural and lithological backgrounds. The time to first instability was delayed by an average of 32.33\%, with a maximum delay of 53.35\%. This effectively improved wellbore stability and control robustness within the studied range of formation parameters.
	\end{enumerate}
	
	In summary, this research provides a systematic framework with potential for engineering application for the stability prediction and intelligent control of drilling in open-hole sections. This can help improve the safety and continuity of complex offshore and extended-reach drilling operations. Nevertheless, the physical fidelity of the current model is still limited by several key idealized assumptions. Future research directions to improve the model’s realism can focus on two aspects: first, introducing anisotropic constitutive relationships that reflect the characteristics of real sedimentary formations into the physical model, and second, coupling the formation and sealing effects of a dynamic wellbore mudcake in the seepage simulation. Integrating these more realistic physical mechanisms into the current intelligent optimization framework is key to further enhancing its predictive accuracy and reliability for field applications.

	\appendix
	
	\section{Experimental Characterization of Rock Mechanical Parameters}
	\label{app:A}
	
\begin{figure*}[t]
		\centering
		\includegraphics[width=\textwidth]{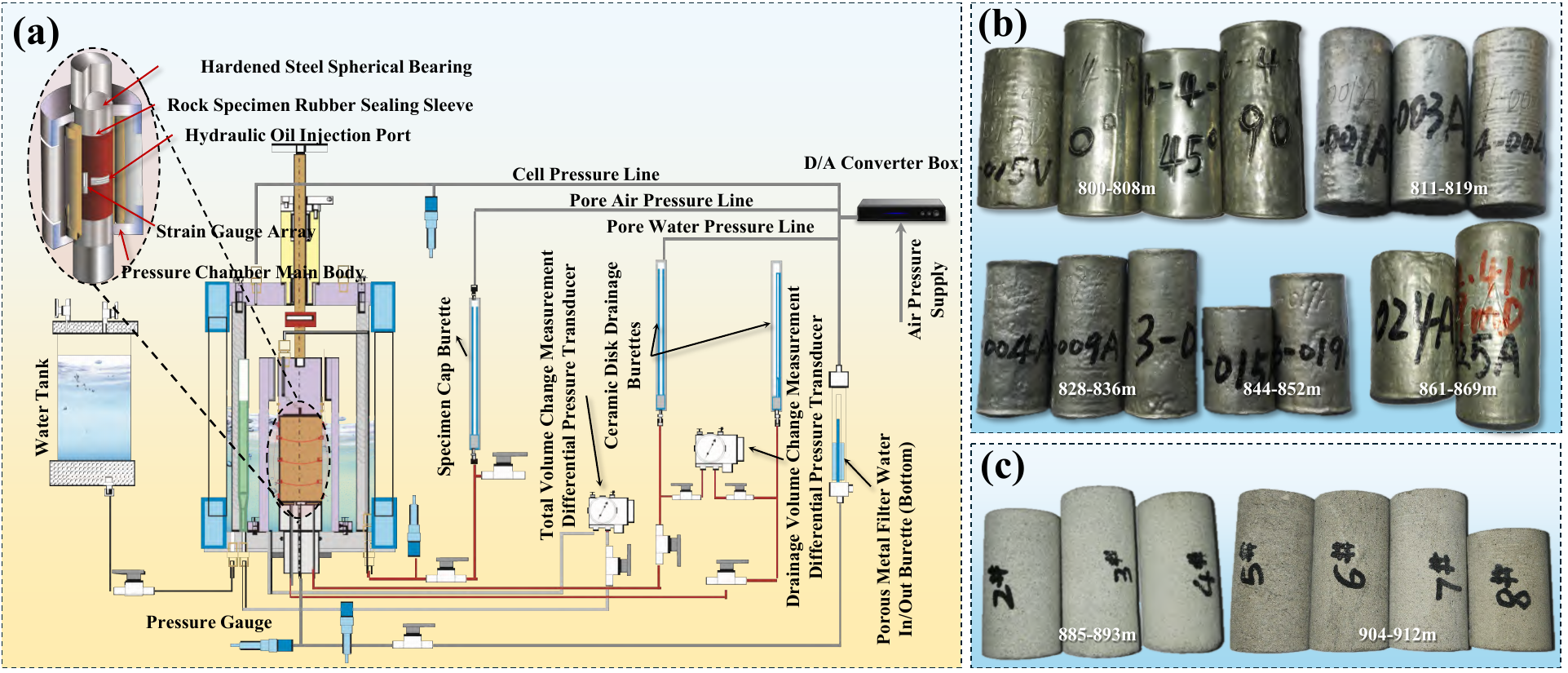}
		\vspace*{-8mm} 
		\caption{Uniaxial/Triaxial Compression Test Apparatus, Jacketed Sample, and Bare Sample.}
		\label{fig:appendix_apparatus}
	\end{figure*}

	To provide high-fidelity geomechanical inputs for the STWM-PINN model, a comprehensive experimental characterization of the formation strength was conducted. This study addresses the high sensitivity of open-hole wellbore stability to the mechanical properties of the rock. A complete strength parameter profile for the 809–910~m target interval was established using uniaxial and triaxial compression tests. These experiments successfully yielded a series of key mechanical parameters, including uniaxial compressive strength, elastic modulus, Poisson’s ratio, cohesion, and internal friction angle. Fig.~\ref{fig:appendix_apparatus} shows the experimental apparatus used in this research and the morphology of typical rock samples.
	
	The preparation of all rock samples strictly followed the procedures of ASTM D7012-14. To ensure specimen homogeneity, the final samples for testing were screened using X-ray Computed Tomography (CT) scanning. Only samples with initial porosity fluctuations within 0.5\% and an anisotropy index below 1.1 were selected. Standard cylindrical specimens, with a diameter of $25.4 \pm 0.5$~mm and a 2:1 length-to-diameter ratio (height $50.8 \pm 0.5$~mm), were obtained by re-coring the original core sections with a diamond hollow drill bit. This aspect ratio was designed to effectively mitigate stress concentration phenomena caused by end constraints. To meet testing requirements, the end faces of each specimen were precision-machined using a dual-face grinder to achieve a surface roughness better than $3.2$~\textmu m and a parallelism tolerance not exceeding $\pm 0.02$~mm. The entire preparation process was conducted under constant temperature ($20 \pm 1$~$^\circ$C) conditions. Low-viscosity kerosene (kinematic viscosity $< 4.1$~mm²/s) was used as a coolant to minimize the generation and propagation of micro-cracks.
	
	The triaxial compression experiments were performed on a high-temperature, high-pressure rock mechanics testing system. The core of the system is a multi-functional confining cell, constructed from forged maraging steel, capable of stable, long-term operation at confining pressures up to 200~MPa and temperatures up to 200~$^\circ$C. Its axial loading unit is a 1000~kN servo-actuator with a displacement resolution better than $0.1$~\textmu m, allowing for continuously adjustable strain rates from $10^{-5}$ to $10^{-2}$~s$^{-1}$. The confining pressure is precisely controlled by an electro-hydraulic servo closed-loop system, which utilizes a plunger-type intensifier to achieve a pressure regulation accuracy of 0.1~MPa. The temperature control unit integrates an annular ceramic heater and a liquid nitrogen cooling circuit, ensuring temperature uniformity across the specimen to within $\pm 1$~$^\circ$C. The experimental procedure followed the recommended methods of the International Society for Rock Mechanics (ISRM). Before loading, 50~\textmu m-thick polytetrafluoroethylene (PTFE) films were placed on the specimen ends to reduce friction. During loading, confining pressure was first applied at a stepwise rate of 0.5~MPa/s to the target value and then held constant for 30~min to ensure complete sample consolidation. Subsequently, axial loading was conducted at a constant strain rate of $10^{-5}$~s$^{-1}$. Throughout the loading process, axial and radial strains were recorded synchronously at a high frequency (100~Hz) to accurately capture post-peak softening and other progressive failure characteristics. To avoid catastrophic failure of the sample, the test was terminated when the principal stress difference dropped to 80\% of its peak strength.
	
	The experimental data were processed based on continuum mechanics theory. The elastic modulus was determined by a least-squares regression on the linear segment of the stress-strain curve. The Poisson’s ratio was then calculated from the ratio of transverse to axial strain. To obtain the strength parameters, peak strength data points from at least three tests under different confining pressures were plotted in Mohr space to construct the ultimate failure envelope. Finally, using the Drucker-Prager criterion consistent with the main physical model, the cohesion and internal friction angle were inverted by fitting this envelope with the Levenberg-Marquardt non-linear least-squares algorithm. This procedure ensured that the obtained parameters could be seamlessly integrated into the subsequent multiphysics coupled model. To construct a complete parameter system, these directly measured mechanical parameters were also supplemented and calibrated with log and test data from offset wells in the block, covering thermo-hydro-chemical parameters such as permeability, thermal conductivity, specific heat capacity, and the hydration diffusion coefficient.
	
	\bibliographystyle{elsarticle-harv} 
	\bibliography{references}
	
\end{document}